\documentclass[aps, singlecolumn, superscriptaddress]{revtex4-2}
\usepackage{xcolor, amsmath, graphicx}
\usepackage{esdiff, relsize, pgfplots, float, graphicx, amsmath, mathtools, ulem}
\begin{document}
	
\title{Tailoring Plasmonics of Au@Ag Nanoparticles by Silica\\ Encapsulation}

\author{Johannes Schultz}
\thanks{J.S. and F.K. contributed equally to this work}
\affiliation{Leibniz Institute for Solid State and Materials Research Dresden, Helmholtzstraße 20, 01069 Dresden, Germany}
\author{Felizitas Kirner}
\thanks{J.S. and F.K. contributed equally to this work}
\affiliation{University of Konstanz, Universitätsstraße 10, 78457 Konstanz}
\author{Pavel Potapov}
\affiliation{Leibniz Institute for Solid State and Materials Research Dresden, Helmholtzstraße 20, 01069 Dresden, Germany}
\author{Bernd Büchner}
\affiliation{Leibniz Institute for Solid State and Materials Research Dresden, Helmholtzstraße 20, 01069 Dresden, Germany}
\affiliation{Institute of Solid State and Materials Physics, Haeckelstraße 3, 01069 Dresden}
\author{Axel Lubk}
\email{a.lubk@ifw-dresden.de}
\affiliation{Leibniz Institute for Solid State and Materials Research Dresden, Helmholtzstraße 20, 01069 Dresden, Germany}
\affiliation{Institute of Solid State and Materials Physics, Haeckelstraße 3, 01069 Dresden}
\author{Elena V. Sturm}
\email{elena.sturm@uni-konstanz.de}
\affiliation{University of Konstanz, Universitätsstraße 10, 78457 Konstanz}

\begin{abstract}
	
	Hybrid metallic nanoparticles encapsulated in oxide shells are currently intensely studied for plasmonic applications in sensing, medicine, catalysis, and photovoltaics. Here, we introduce a method for the synthesis of Au@Ag@SiO$_2$ cubes with a uniform silica shell of variable and adjustable thickness in the nanometer range; and we demonstrate their excellent, highly reproducible, and tunable optical response. Varying the silica shell thickness, we could tune the excitation energies of the single nanoparticle plasmon modes in a broad spectral range between 2.55 and 3.25\,eV. Most importantly, we reveal a strong coherent coupling of the surface plasmons at the silver-silica interface with the whispering gallery resonance at the silica-vacuum interface leading to a significant field enhancement at the encapsulated nanoparticle surface in the range of 100\,\% at shell thicknesses $t\,$$\simeq\,$20\,nm. Consequently, the synthesis method and the field enhancement open pathways to a widespread use of silver nanoparticles in plasmonic applications including photonic crystals and may be transferred to other non-precious metals.
	
\end{abstract}

\maketitle
	
	\section{Introduction}
	\label{sec:intro}
	Localized surface plasmon (LSP) resonances of (noble) metal nanoparticles (NPs) with a size comparable or smaller than the wavelength of incident light lead to intriguing optical properties such as frequency-dependent electromagnetic field confinement and amplification. These are exploited in a multitude of applications, such as sensors\cite{Anker2008}, spectroscopy signal enhancement\cite{He2017}, nanoantennas\cite{Goris2014} and catalysis\cite{Garcia-Cruz2019}, where the frequency of the surface plasmons on individual particles can be tuned specifically by varying compositions (e.g., use of different metals or alloys)\cite{Schletz2021, Crespo2014}, morphology and size of the particles\cite{Burda2005}, and environment (e.g., different dielectric surroundings)\cite{Hanske2018}. The latter offers a particular rich playground ranging from mere frequency tuning over generation of hybridized surface plasmon modes\cite{Vernon2010, Chen2012, Prodan2003} and plasmon bands\cite{Mayer2019, Schletz2021} in nanoparticle oligomers, polymers, and superlattices, to strong coherent coupling between surface plasmons and other excitations such as excitons\cite{Hanske2018}. Here, the NP morphology, their interparticle distance, and arrangement within the assembly define the coupling strength and hence the resonance energies of the coupled mode.\cite{Nordlander2004, McMahon2011, Halas2011, Jain2007, Hooshmand2019, Mayer2019}\\
	Generally, the fabrication of plasmonic nanostructures may be categorized into top-down and bottom-up approaches. Electron-beam lithography\cite{Jain2007, Zhu2011, Menumerov2018} belongs to the latter and allows a precise control of dimension, geometry, and interparticle distance of the NPs. Zhu et al.\cite{Zhu2011}, for instance, present a method to lithographically produce pairs of silver particles with narrow 3\,nm gaps. Notwithstanding, this pathway also faces several downsides concerning the crystal quality of lithographically fabricated (noble) metal nanostructures, control over 3D shape of nanoparticles down to nanometer length scales\cite{Zhu2011, Menumerov2018} and fabrication of large NP arrangements in particular 3D photonic crystals\cite{Sun2018}.\\
	A bottom-up approach for the design of nanostructures that partially overcomes the aforementioned issues is the (self-)assembly of NPs. A particularly simple method is the drying of a dilute NP dispersion.\cite{Gao2012} More sophisticated methods include DNA origami\cite{Roller2017, Zhao2020}, assembly by a gas-phase diffusion technique\cite{Bahrig2014, Brunner2020} or under weak external magnetic fields\cite{Kapuscinski2020}, and self-assembly on patterned substrates\cite{Mayer2019}. The advantage of these methods lies in the high quality of the NPs and the straightforward upscaling of some of the self-assembly methods. Drawback is the reduced flexibility in deliberately arranging the NPs compared to lithographical methods, e.g., the precise adjustment of interparticle distances\cite{Hanske2018} or the deliberate creation of NP superlattices (e.g. colloidal crystals, mesocrystals\cite{Sturm2016}). However, to purposefully influence the superlattice arrangement is out of the scope of this paper and needs additional elaboration.\\
	Highly crystalline noble metal NPs with specific shapes and sizes can be synthesized by a variety of methods. One of the most prominent procedures for the synthesis of well-defined particles is the seed-mediated growth method.\cite{Jana2001, Niu2013} Therein, nucleation and growth process are largely separated, which allows a individual optimization of experimental parameters. NPs are thermodynamically unstable due to their high surface to volume ratio which is why most syntheses employ surfactants to stabilize the NPs and control their morphology.\cite{Heinz2017, Kirner2020, Grzelczak2008, Boles2016} Single crystalline gold NPs can be reproducibly synthesized in aqueous solution at mild reaction conditions.\cite{Zheng2013} Silver NPs may require more elaborate methods with harsher conditions and the reproducibility depends on several synthesis parameters.\cite{Jeon2014} Yet, gold-silver core-shell nanocubes can be produced at high yields using a seed-mediated synthesis with single crystalline gold particles as seeds and cetyltrimethylammonuim chloride (CTAC) as surfactant.\cite{Gomez-Grana2013} This epitaxial overgrowth of silver onto gold is a less elaborate process leading to a higher reproducibility compared to the synthesis of single crystalline silver NPs.\\
	It is important to note that surfactants often deteriorate the plasmonic properties of metallic NPs by modifying the dielectric properties of the crucial metal-to-environment interfaces. After the assembly of NPs, remaining surfactants should be therefore removed from the particle surface, e.g., by plasma cleaning. However, a complete or nearly complete removal of the stabilizer can lead to destabilization, shape-loss, and fusion of the NPs. In particular, unprotected silver NPs are quite sensitive to modification and deterioration. For instance, the oxidization of silver can cause distorted NPs with a silver-oxide shell of poor optical response. The shell's thickness is mostly unclear and changing.\cite{Hanske2018} In consequence, the plasmonic response of silver NPs deteriorates, shifts over time and is generally not-well defined.\cite{Schletz2021}. This is one of the reasons why silver is a relatively unpopular plasmonic material, despite its broad plasmon band and low loss in the optical regime.\\
	These stability issues can be circumvented with an encapsulation by silica (high optical transmittance in the optical frequency range) as shown by flame aerosol method for the synthesis of Au@Ag NPs followed by a swirl injection of silica precursor vapor.\cite{Sotiriou2010} For NPs in solution, a silica encapsulation is generally performed using a modified Stöber process\cite{Stober1968, Kobayashi2005} following the NP synthesis and removal of excess surfactants (e.g., CTAC). As a transfer of the NPs from aqueous solution to an ethanol-water mixture is necessary, the high surface potential of positively-coated NPs needs to be screened first by polyelectrolytes or polymers to avoid aggregation.\cite{Pastoriza-Santos2006} As a result, after encapsulation stable NPs with a well-defined metal-silica and silica-environment (e.g., air) interface are obtained. Such encapsulation of NPs can be also exploited as an additional degree of freedom for tuning plasmon modes as both the size of the NPs\cite{Hung2013} and the thickness of the silica shell\cite{Rodriguez-Fernandez2007, Kluczyk-Korch2019} influence the excitation energy of the LSP modes. Moreover, by adjusting the thickness of the silica coating, a flexible adaption of the interparticle distance in assemblies is possible. Consequently, silica encapsulation represents a viable route for largely improving plasmonic properties of silver and other metallic NPs and their assemblies, greatly facilitating their use in plasmonic applications.\cite{Hanske2018}\\
	In this study, we present an approach to effectively replace the CTAC-capping of Au@Ag NPs with a silica-shell of uniform and adaptable thickness, whereby the thickness of the silica shell can be controlled with high precision. Highly spatially resolved plasmon excitation maps obtained from Electron Energy-Loss Spectrum imaging in a Scanning Transmission Electron Microscope (STEM-EELS) demonstrate an excellent and highly reproducible optical response of the encapsulated NPs, which may be tuned through the shell thickness. Most notably, we reveal a strong enhancement of the plasmonic fields in the encapsulated NPs mediated through a coherent coupling of the surface plasmons of the Au@Ag NPs and the Mie resonances of the silica shell. We also demonstrate that this effect leads to an enhanced hybridization of coupled NPs which opens new possibilities for plasmonic applications.
	
	\section{Results and Discussion}
	\label{sec:results}
\begin{figure}[h!]
		\centering
		\includegraphics[width=\textwidth]{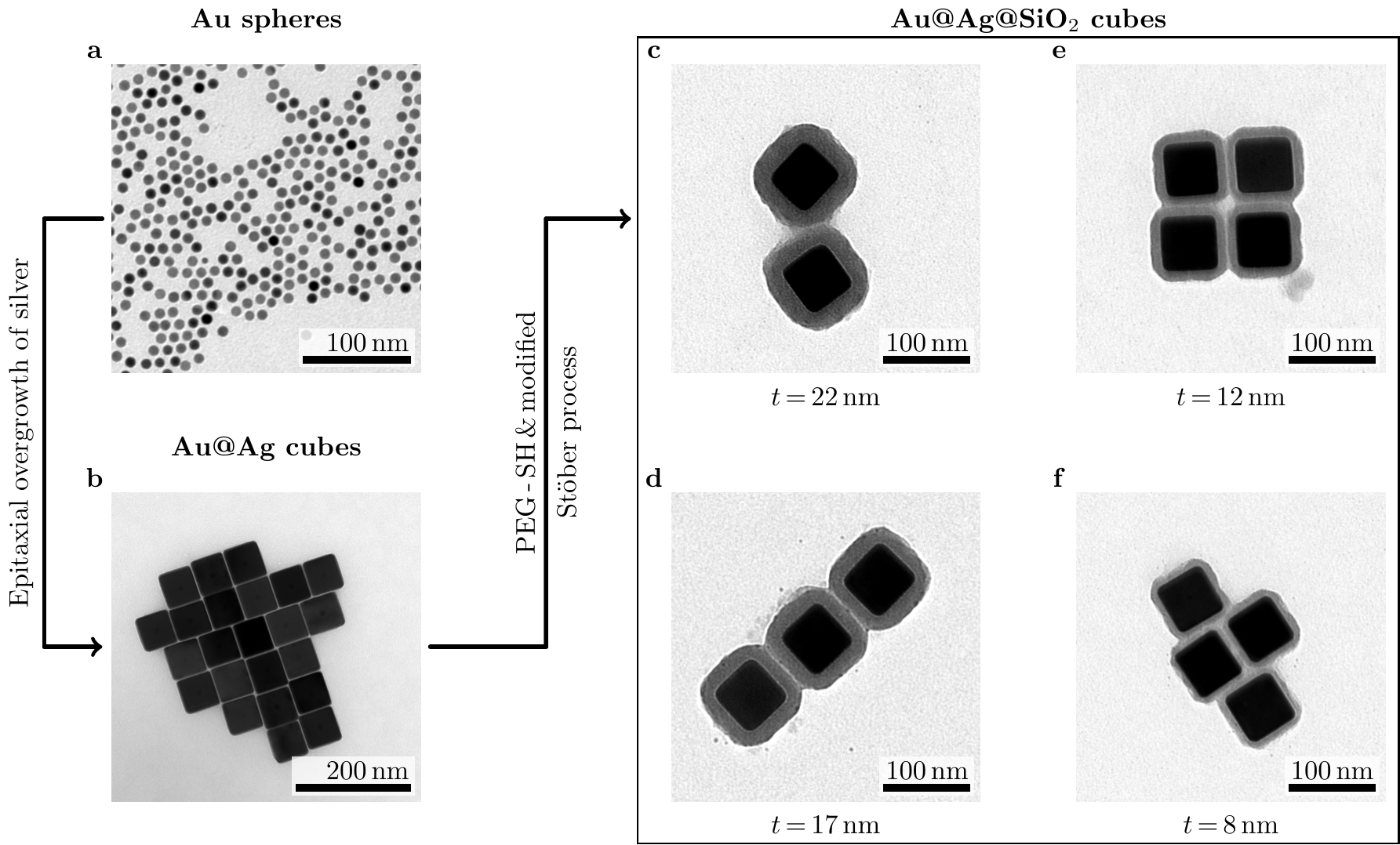}
		\caption{Schematic synthesis procedure for silica-encapsulated Au@Ag core shell particles. First, gold spheres are synthesized (a) that, second, are overgrown with silver (b). After functionalization with PEG-SH they are encapsulated by a silica-shell (c-f) using a modified Stöber process.}
		\label{fig:synthesis}
	\end{figure}

\begin{figure}
		\centering
		\includegraphics[width=\textwidth]{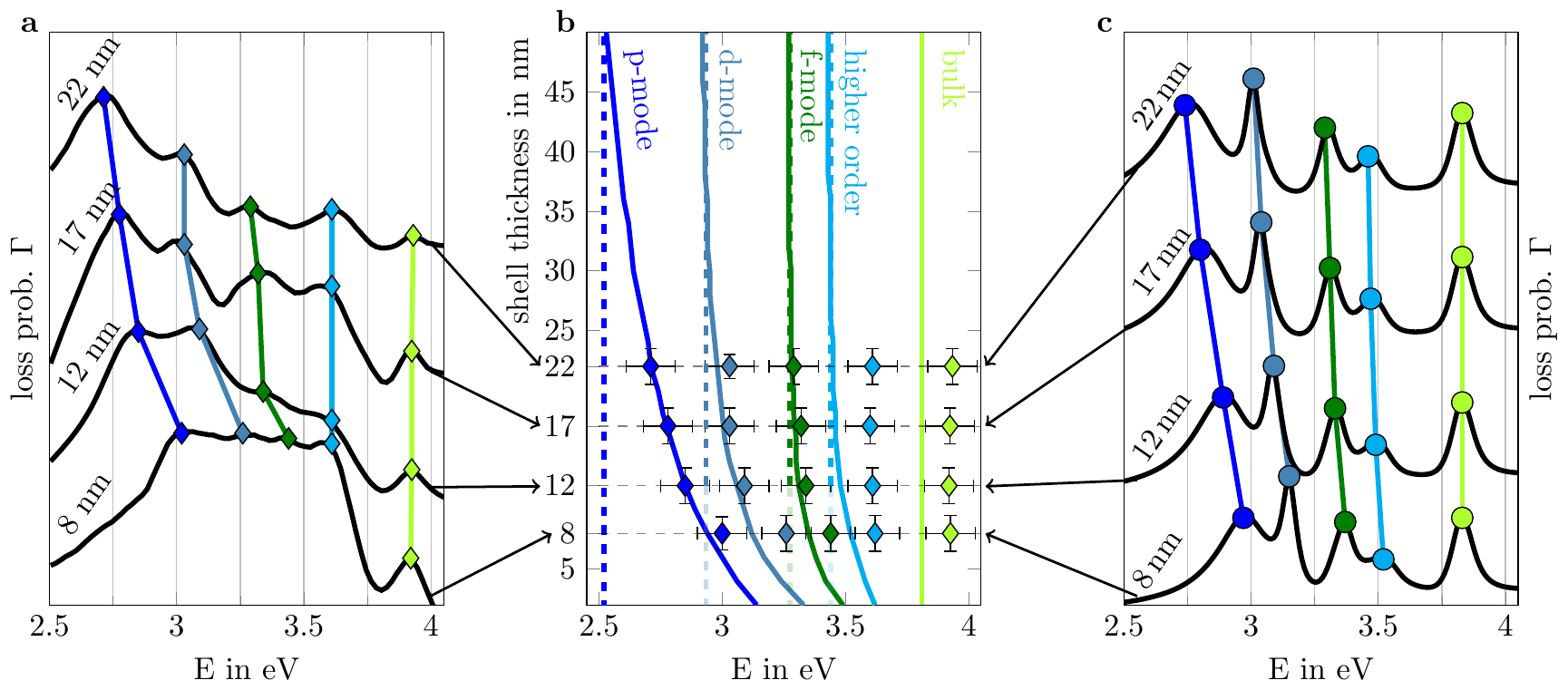}
		\caption{Experimental (a) and simulated (c) EEL spectra of 60\,nm Au@Ag cubes with varying silica shell thickness. (b) Represents the comparison of experimental (diamonds) and simulated (solid line) LSP mode positions including mode character and asymptotic limits (dashed lines).}
		\label{fig:spectra}
	\end{figure}

\begin{figure}[h!]
		\centering
		\includegraphics[width=\textwidth]{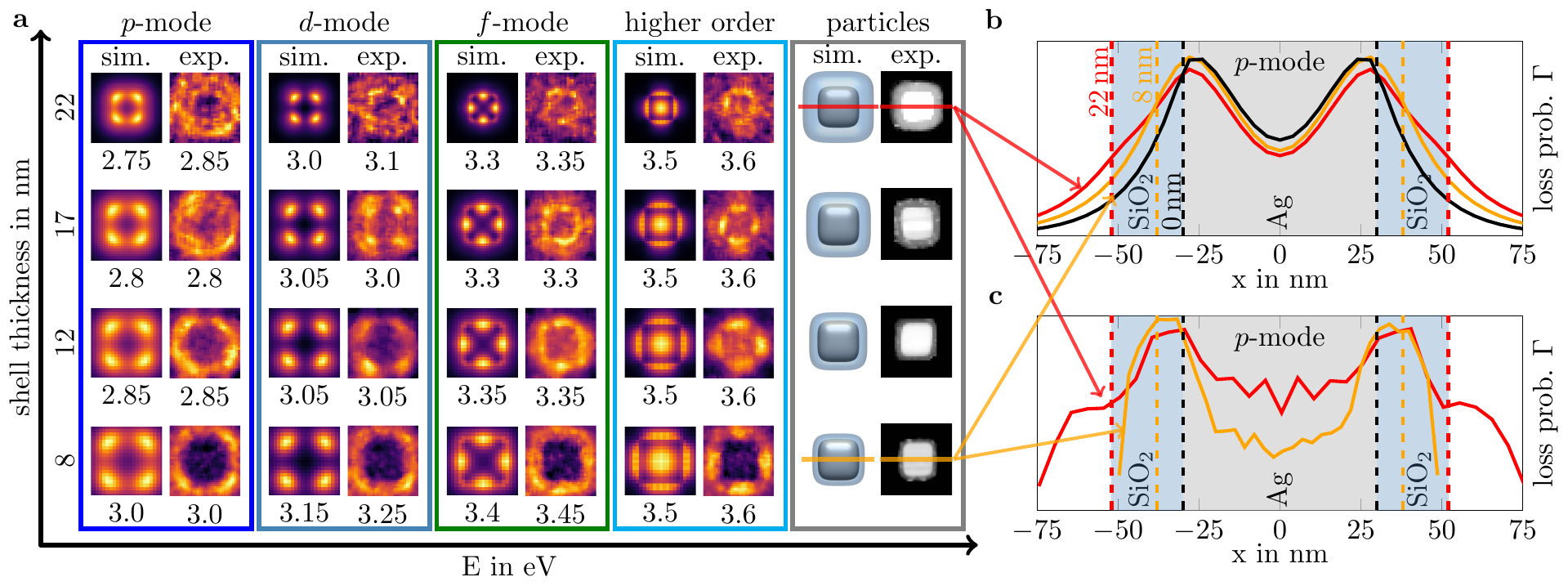}
		\caption{(a) Simulated and experimental EEL probability maps in dependence of the silica shell thickness. (b) Simulated and (c) experimental loss probability line plots of the p-mode for 22\,nm (red line) and 8\,nm (orange line) silica shell thickness, respectively. Note the enhancement of the loss probability at the silica-vacuum interface (orange/red dashed lines respectively).}
		\label{fig:maps}
	\end{figure}

\begin{figure}[h!]
		\centering
		\includegraphics[width=\textwidth]{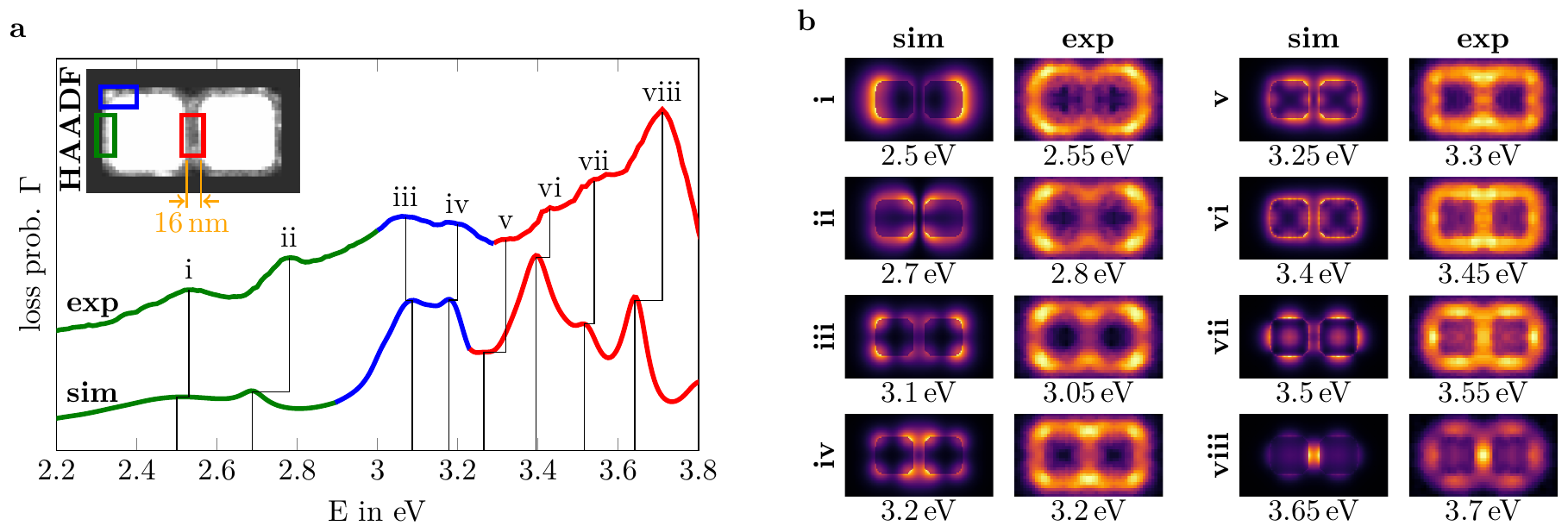}
		\caption{(a) Experimental and simulated EEL spectra for a Au@Ag-cube dimer coated with 8\,nm silica. Here, the color of the curve corresponds to different spatial sub-areas of the EELS scan, over which the loss probability is averaged. The different sub-areas are indicated by the red, blue, and green rectangles in the HAADF image of the dimer. (b) EEL probability maps of the modes indicated (i - viii) in (a) respectively. The enhanced coupling of the LSP modes mediated by the silica coating results in a hybridization of the modes excitable in the single NPs respectively. Note, that the experimental obtained EEL probability maps are symmetrized with respect to the central plane of the dimer in order to enhance the signal to noise ratio.}
		\label{fig:dimer}
	\end{figure}

	The Au@Ag nanocubes (Figure\,\ref{fig:synthesis}(a),\,(b)) are synthesized as previously published by G\'{o}mez-Gra\~{n}a et al.\cite{Gomez-Grana2013} and described in the experimental section. Au@Ag nanocubes with an edge length of (60$\,\pm\,$1)\,nm are used for further modification. The growth of the silica layer is achieved by a modified procedure for gold NPs published by Monta\~{n}o-Priede et al.\cite{Montano-Priede2017} The overgrowth of the NPs with silica is conducted by a modified Stöber process\cite{Stober1968} and requires a transfer of the Au@Ag NPs from aqueous solution to a solution of ammonia in ethanol (Figure\,\ref{fig:synthesis}(c)-(f)). The positive charge of the CTAC-capped NPs needs to be screened before the transfer to avoid their aggregation and flocculation.\cite{Pastoriza-Santos2006} This was achieved by washing and absorption of $\omega$-thiol-terminated polyethylene glycol (PEG-SH) on the particle. Zeta potential measurements prove the change of the surface potential upon surfactant exchange (Figure\,S5). The thickness of the silica shell can be tuned by varying the amount of tetraethyl orthosilicate (TEOS) added to the reaction solution which leads to a very precise control of surface plasmon energies in the following. Note, however, that the fabrication of silica shells with a thickness below a few nanometer, remains challenging. In order to systematically study the influence of the silica shell, Au@Ag NPs with an edge length of 60\,nm and silica layers of 8, 12, 17, and 22\,nm (see Figure\,\ref{fig:synthesis}) were characterized by spatially and spectrally highly-resolved STEM-EELS experiments (see experimental section for details).\\
	The EEL spectra of the cubic NPs exhibit five peaks corresponding to four different LSP modes and the silver bulk plasmon peak (Figure\,\ref{fig:spectra}). The experimental spectra (Figure\,\ref{fig:spectra}\,(a)) are in striking agreement (Figure\,\ref{fig:spectra}\,(b)) with the simulated results (Figure\,\ref{fig:spectra}\,(c)), which demonstrates the high quality of the silica shells and the dielectrically well-defined silver-silica interface. Here, the influence of the small gold nucleus in the Au@Ag NPs can be safely neglected due to the confinement of the LSPs to the silver-silica interface. Furthermore, the slightly increased broadening of all experimental peaks including the bulk plasmon is ascribed to imperfections of the silver structure of the NPs. In accordance with previous studies, the asymptotically converging red shift of the LSP modes increases with the silica shell thickness.\cite{Rodriguez-Fernandez2007, Kluczyk-Korch2019} The latter opens pathways to finely tune the optical response of the NPs.\\
	In order to classify the LSP modes, we employ a cubic harmonic basis derived from the angular momentum classification ($l$$\,=\,$$1\,\widehat{=}\,$p, $l$$\,=\,$$2\,\widehat{=}\,$d, $l$$\,=\,$$3\,\widehat{=}\,$f, …) of LSP eigenmodes in close agreement with atomic orbital categorization (see appendix\,\ref{sec:S4}). The LSP mode with highest excitation energy (3.65\,eV at bare 60\,nm Au@Ag nanocubes), denoted as fundamental mode in the following, corresponds to a superposition of energetically almost degenerated modes of high order in $l$. For small values of $l$ the excitation energies of the LSP modes differ noticeable from the fundamental mode whereby the energy splitting increases with decreasing value of $l$. In case of bare Au@Ag nanocubes, the p, d and f mode corresponding to 3.25, 3.45 and 3.6\,eV, respectively, are clearly discernible from the fundamental mode at 3.65\,eV (Figure\,\ref{fig:spectra}). Accordingly, the LSP mode with the lowest excitation energy (p-mode) can be tuned in a range between 2.55 and 3.25\,eV by increasing the silica shell thickness. On the other hand, the tunable range of the higher order LSP mode is very narrow (0.23\,eV).\\
	The observed red shift can be understood by describing the system with an adapted dielectric media approach. Here, the complex dielectric surrounding is described by an effective dielectric function obtained from a weighted average of silica and vacuum dielectric functions depending on the skin depth $d_s=|k^{-1}|$ of the plasmon wave that is exponential damped away from the interface ($\propto$$\,e^{-kz}$, where $z$ is the distance from the interface and $k$ is the plasmon wavevector). The logic behind this approach is as follows: when the thickness of the silica shell approaches zero, the effective dielectric function needs to match the dielectric function of vacuum. On the other hand, for the infinitely thick silica shell, the effective dielectric function should be equal to the dielectric function of silica. Between the two limits, the effective dielectric function follows a monotonous course, which is determined from the skin depth dependent weighting of the dielectric functions. For an infinite half-plane, the skin depth can be calculated analytically (see appendix \ref{sec:S5}). For a silver-silica interface, the skin depth increases with decreasing excitation energy (see Figure\,S7\,(b)). Therefore, the tunable range of excitation energy is larger for modes with lower excitation energy.\\
	Similar to the resonance energy positions shown in Figure\,\ref{fig:spectra}, the spatially resolved loss probability maps shown in Figure\,\ref{fig:maps}\,(a) excellently match the theoretical predictions. Minute spatial details of the SPR excitation are reproduced on length scales down to 5\,nm. This again corroborates the high-quality of both the silica shell and the Au@Ag NP and their stability under both the electron beam and ambient conditions (see, e.g., Ref.\cite{Schletz2021} for degraded optical properties of bare Au@Ag NPs).\\
	Note furthermore that both the experimental and simulated spatially resolved surface plasmon maps reveal another intriguing feature of the thin silica shells, a strongly enhanced loss probability at the silica-vacuum interface compared to the bare Au@Ag NP reference in the range of 100\,\% (see 1D cross section in Figure\,\ref{fig:maps}\,(b),\,(c). Considering that the loss probability is proportional to the $z$-component of the induced electric field of the LSP, this strong field enhancement directly increases the interaction of the silica-encapsulated NPs with their surroundings, e.g., molecules attached to the shell (e.g., in sensing applications) or other NPs (e.g., in photonic crystals, see below).
	The effect arises from a resonant coupling between the surface plasmon modes located at the silica-silver interface and a broad Mie type resonance located at the vacuum-silica interface. As the vacuum-silica interface exhibits a drop of the dielectric function from 2 to 1 in the considered energy range the latter resonance can may be associated with a whispering gallery type mode. The rather broad spectral line, not typical for whispering gallery modes, can be attributed to the non-spherical shape of the nanocubes and their small size (see appendix\,\ref{sec:S6}), extending the field enhancement effect over a broad energy range.\\
	To further corroborate the crucial field enhancement effect, we consider a simple coupling scenario, i.e., the formation of hybridized plasmon modes in a dimer configuration (see Figure\,\ref{fig:dimer}). It is well-known that the formation of split energy hybridized plasmon modes from (degenerated) single particle LSPs crucially depends on the coupling strength between the NPs. Accordingly, the coupling strength may be read-off from the energy splitting of the single particle modes. It quickly decays upon increasing interparticle distances beyond several nanometers in case of the bare 60\,nm Au@Ag NPs, leading to a barely noticeable coupling effect if the interparticle distance exceeds 48\,nm (see appendix\,\ref{sec:S7}). On the other hand, the field enhancement due to the silica shell strongly increases the interaction of the single particle modes (see appendix\,\ref{sec:S7}), leading to significantly stronger coupling effects (i.e., formation of distinct hybridized plasmon modes) up to 100\,nm interparticle distance (50\,nm silica shell). The hybridization leads to split energy levels of the LSP modes in the dimer EEL spectrum, e.g., the modes indicated by i and ii in Figure\,\ref{fig:dimer}\,(a) arising from hybridization of the single-particle $p$-mode (see Figure S7 for a detailed explanation of the the mode classification). The hybridized nature of the LSP modes is also visible through the spatially resolved loss maps plotted in Figure\,\ref{fig:dimer}\,(b). Both, the dimer EEL spectrum as well as the loss probability maps are again in striking agreement between experiment and simulation. We can conclude that the field enhancement is clearly expressed in the hybridized modes of the dimer.
	\newpage
	\section{Conclusion}
	\label{sec:conclusion}
	In summary, we transferred the fabrication process of silica shell encapsulation from gold to Au@Ag nanoparticles and thereby protected the silver surface. Employing high-resolution electron energy loss spectroscopy we demonstrate very well-defined dielectric response of the encapsulated Au@Ag nanoparticles, not affected by environmental degradation (most notably oxidation). That includes reproduction of minute spatial details of the localized surface plasmon excitation on length scales down to 5\,nm. By modulating the silica shell thickness with nanometer precision single particle surface plasmon excitation energies could be tuned in a wide energy range between 2.55 and 3.25\,eV. This behavior could be explained through an adapted dielectric media approach. Most notably, we discovered and explained a strong field enhancement at the vacuum-silica interface of the silica encapsulated Au@Ag nanoparticles, originating from resonant coupling between localized surface plasmons at the silica-silver interface and a Mie type resonance at the vacuum-silica interface. We demonstrate the magnitude of the effect by creating strongly hybridized localized surface plasmon modes in a nanoparticle dimer of 16\,nm interparticle distance. Our findings open pathways towards fabrication of plasmonic oligomers, polymers and colloidal crystals of well-defined geometry and structure (incl. interparticle distance), stability against environmental degradation, and strongly hybridized plasmon modes and bands over a large frequency range; and hence applications of the latter in sensing and signal transfer.

	\section{Experimental Section}
	\label{sec:experimental}
	\subsection{Chemicals}
	Milli-Q water with a resistivity of 18.2\,MΩ$\,\cdot\,$cm was used in all experiments. Chemicals were bought and used as received. Hexadecyltrimethylammonium bromide (CTAB, $\geq$ 99\,\%), hydrogen tetrachloroaurate trihydrate (HAuCl$_4$\,$\cdot$\,3\,H$_2$O, $\geq$ 99.9 \,\%), cetyltrimethylammonium chloride solution (CTAC, 25 wt.\,\% in H$_2$O) and tetraethyl orthosilicate (TEOS, for synthesis) were purchased from Sigma-Aldrich. Sodium borohydride (NaBH$_4$, $\geq$ 97\,\%), L(+)-ascorbic acid (AA, $\geq$ 99\,\%,), silver nitrate (AgNO$_3$, $\geq$ 99\,\%) and ethanol (EtOH, $\geq$ 99,5\,\%, Ph.Eur., extra pure) were purchased from Roth. $\alpha$-Methoxy-$\omega$-mercapto polyethylene glycol (CH$_3$O-PEG-SH, M$_{\text{w}}$ 5000 Dalton) was purchased from Rapp Polymere. Ammonia solution (NH$_4$OH, aq. 25\,\%, for analysis) was purchased from Merck.\\
	\medskip
	\subsection{Characterization}
	\textbf{UV/Vis characterization.} UV/Vis spectra were recorded using a Varian Cary 50 UV/Vis Spectrophotometer.\\
	\textbf{Zeta potential measurements.} Zeta potential was measured on a Malvern Zetasizer Nano ZSP at 25\,°C with 5$\cdot$12 runs at a count rate of 197.8\,kcps.\\ 
	\textbf{AUC measurements.} The analytical ultracentrifugation (AUC) measurements were performed on an Optima XL I (Beckman Coulter, Palo Alto, CA, United States) using Rayleigh interference optics and 12\,mm double sector titanium centerpieces (Nanolytics, Potsdam, Germany). A SW 60 Ti Swinging-Bucket Rotor was used at 3000\,RPM and 20\,°C.\\
	\textbf{STEM EELS characterization.} To characterize the plasmonic response of the silica coated Au@Ag NPs Scanning Transmission Electron Microscopy in combination with Electron Energy-Loss Spectroscopy\cite{Nelayah2007,Schmidt2012,Martin2014,Yoshimoto2018} was applied in a probe Cs-corrected FEI Titan$^3$ Transmission Electron Microscope (TEM). A focused electron-beam was scanned over the NPs with a dwell time of 30\,ms. At each scanning point an EEL spectrum was recorded using a Gatan Tridiem energy filter. Simultaneously high angular dark field (HAADF) images under the convergence angle of 20\,mrad and a collection angle of 7\,mrad were acquired. The primary energy of 80\,keV of the beam electrons was monochromized to $\pm$40\,meV using the Wien filter of the TEM. To avoid contamination, the samples were plasma-cleaned for 8\,s in advance. Using this technique, the position of the characteristic peaks in the EEL spectra corresponding to LSP modes excited by the beam electrons were studied for NPs with different silica shell thickness. Furthermore, the spatial distribution of the induced plasmonic fields were studied by extracting spatially resolved EEL probability maps of the different LSP modes. To correct for the scattering absorption, the spatial resolved EELS signal was normalized by the overall intensity at the corresponding scanning position respectively ($\Gamma_{\text{norm}}(x,y,\omega)=\Gamma(x,y,\omega) / \int \Gamma(x,y,\omega) \text{d}\omega$).\\
	\medskip
	\subsection{Synthesis of Nanoparticles}
	\textbf{Synthesis of initial gold seeds.} An aqueous solution of HAuCl$_4$ (0.25\,mL, 0.01\,M) and an aqueous solution of CTAB (7.50\,mL, 0.10\,M) were mixed in a glass vial (50\,mL) and tempered to 27\,°C. Freshly prepared, ice-cold NaBH$_4$ solution (0.60\,mL of 0.01\,M) was added under vigorous stirring. The seed solution was aged for 90\,min at 27\,°C to ensure complete decomposition of excess borohydride.\\
	\textbf{Synthesis of spherical gold seeds.} Spherical gold seeds with a diameter of 9.5\,nm were synthesized as previously described.\cite{Kirner2020, Zheng2013} In a typical synthesis, an aqueous solution containing CTAC (39.00\,mL, 0.10\,M) and HAuCl$_4$ (1.00\,mL, 0.01\,M) was tempered to 27\,°C in a water bath. AA (15.00\,mL, 0.10\,M) was added, followed by a rapid injection of 500$\,\mu$L initial seeds. The solution was kept at 27\,°C for 15\,min. The spherical seeds were characterized by UV/Vis spectroscopy, TEM and AUC. The polydispersity index (PDI) of the samples was 1.04 (Figure\,S1-S2, Equation\,S1).\\
	\textbf{Synthesis of Au@Ag cubes.} Au@Ag cubes were synthesized by a procedure published by G\'{o}mez-Gra\~{n}a et al.\cite{Gomez-Grana2013} Typically, water (21.00\,mL), CTAC (3.00\,mL, 0.10\,M), AA (1.20\,mL, 0.10\,M) and 275$\,\mu$L of as prepared spherical seeds were mixed. Then, AgNO$_3$ (3.00\,mL, 0.01\,M) was added and the solution was heated to 65\,°C and left undisturbed for 12\,h.\cite{Gomez-Grana2013} The particles were collected by centrifugation (5000\,RPM, 5\,min). The Au@Ag cubes were characterized by TEM and UV/Vis spectroscopy (Figure\,S3). The yield of cubic Au@Ag particles was 85\,\% (Figure\,S4).\\
	\textbf{Synthesis of Au@Ag@SiO$_2$ cubes.} The synthesis was adapted from previously published procedure for the silica coating of gold NPs by Monta\~{n}o-Priede et al.\cite{Montano-Priede2017} 60\,mL of the synthesized Au@Ag cubes were centrifuged at 9000\,RPM for 4\,min. The supernatant was removed, and the particles were redispersed in CTAC (28.00\,mL, 0.05\,M). The dispersion was centrifuged at 6000\,RPM for 10\,min. The supernatant was removed, and the particles were redispersed in 5\,mL MilliQ containing 0.79\,mg PEG-SH, afterwards CTAC (5.00\,mL, 0.05\,M) was added. The solution was left undisturbed for 1\,h. The particles are concentrated to 1.35\,mL at 6000\,RPM for 10\,min. Two further solutions were prepared. For the first solution (solution 1), 1\,mL NH$_4$OH was added to 37.5\,mL of EtOH. For the second solution (solution 2), 50$\,\mu$L TEOS were added to 7.4\,mL of EtOH. The concentrated NP solution (1.35\,mL) is added to 2.5\,mL EtOH and 3.85\,mL of solution 1. Then, varying volumes of solution 2 are added (Table S1). The reaction mixture was stored at 30\,°C for 2\,h. For the preparation of TEM samples 750$\,\mu$L of the dispersion and 150$\,\mu$L of 5\,mM CTAC are centrifuged at 6500\,RPM for 5\,min and redispersed in 1\,mM CTAC solution. Zeta potential measurements of particles capped with CTAC and PEG-SH are depicted in Figure\,S5.\\
	\medskip
	\subsection{Electromagnetic simulations}
	The plasmonic response of the NPs was simulated using a Boundary Element Method (MNPBEM code) implemented by Hohenester et al.\cite{Hohenester2012, Hohenester2014}. Applying the MNPBEM code the plasmonic response was calculated by solving the full Maxwell equations for 80\,keV electrons exciting plasmon modes in silica coated silver cubes with 60\,nm edge length. The dielectric functions of silver and silica are taken from Ref.\cite{Yang2015} and \cite{Gao2012a}. The medium surrounding the NPs was assumed to be vacuum ($\epsilon$\,=\,1). The influence of the small gold nucleus in the Au@Ag NPs can be safely neglected in the simulation due to the confinement of the plasmons to the silver-silica interface.

	\medskip
	\textbf{Acknowledgements} \par
	Johannes Schultz and Felizitas Kirner contributed equally to this work. A.L. has received funding from the European Research Council (ERC) under the Horizon 2020 research and innovation program of the European Union (grant agreement no. 715620). J.S. received funding from the Deutsche Forschungsgemeinschaft (DFG, German Research Foundation) under Germany's Excellence Strategy through Würzburg-Dresden Cluster of Excellence on Complexity and Topology in Quantum Matter ‐ ct.qmat (EXC 2147, project‐id 390858490). F.K and E.S. acknowledge Deutsche Forschungsgemeinschaft (DFG, German Research Foundation) under SFB 1214, project B1 and Zukunftskolleg at the University of Konstanz for financial support. P.P. received funding from the Deutsche Forschungsgemeinschaft (project-id 431448015). The authors thank Rose Rosenberg for AUC measurements and data evaluation.

	\newpage
	\appendix
		\section{Spherical Gold Seed Nanoparticles}
		\label{sec:S1}
		The Au@Ag@SiO$_2$ nanoparticles were synthesized in a seed mediated, multi-step procedure. In the first step, spherical gold seeds were synthesized. The quality of the particles was investigated by Transmission Electron Microscopy (TEM) and Analytical Ultracentrifugation (AUC) (Figure\,\ref{fig:S1}). As TEM is prone to be biased towards larger sizes due to sample preparation, spheres were additionally investigated with AUC. Size measurements from TEM and AUC can slightly differ due to influence of the surfactant shell on the hydrodynamic radius in AUC. As the surfactant shell is not important for the further synthesis, the core diameter was determined from TEM, while the narrow size distribution was confirmed by AUC. The polydispersity index (PDI) of the sample was 1.04 and was calculated as the ratio of the weighted average diameter $D_w$ to the number average diameter $D_n$ (Equation\,\ref{eq:S1}-\ref{eq:S3}, here $x_i$ corresponds to the $i^{\text{th}}$ sample and $w_i$ to the $i^{\text{th}}$ weight).
		\renewcommand{\thefigure}{S1}
		\begin{figure}[h!]
			\centering
			\includegraphics[width=\textwidth]{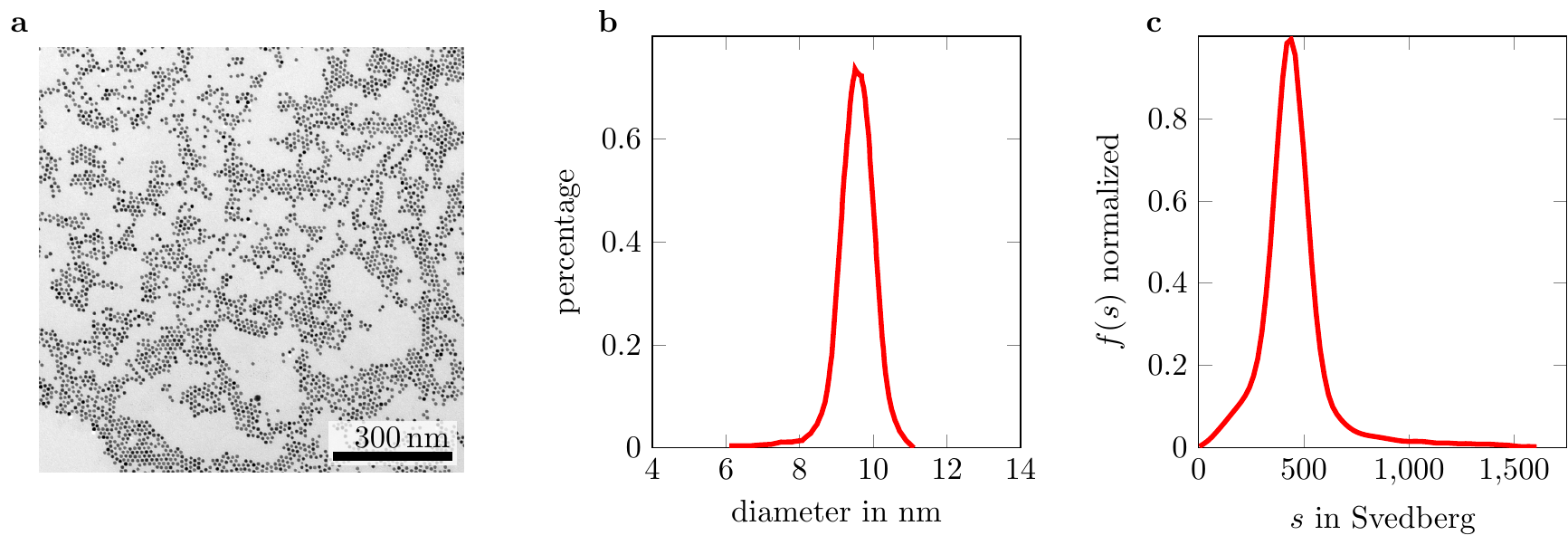}
			\caption{The diameter of the spherical gold seeds was investigated by means of TEM and AUC. (a) Bright-field TEM image of the spherical gold seeds. The median diameter was evaluated from bright-field TEM images using Fiji\cite{Schindelin2012} yielding 9.5\,nm. (b) The size distribution was investigated in AUC. (c) The size is based on sedimentation velocity analysis of the sample.}
			\label{fig:S1}
		\end{figure}
		\renewcommand{\thefigure}{S2}
		\begin{figure}[h!]
			\centering
			\centering
			\includegraphics[trim={5.2cm 0 0 0}, clip]{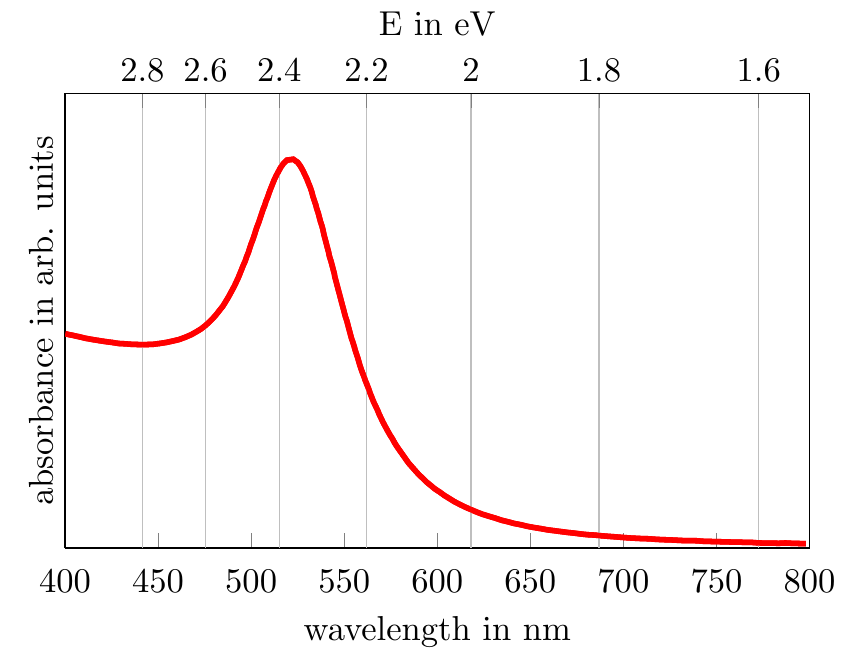}
			\caption{UV/Vis absorbance spectrum of the spherical gold seeds. The absorbance maximum is at 522\,nm.}
			\label{fig:S2}
		\end{figure}
		\newpage
		\begin{equation}
			PDI=\frac{D_w}{D_n}
			\tag{S1}
			\label{eq:S1}
		\end{equation}
		\begin{equation}
			D_w=\frac{\sum_{i=1}^{n} w_ix_{i}^2}{\sum_{i=1}^{n} w_ix_i}
			\tag{S2}
			\label{eq:S2}
		\end{equation}
		\begin{equation}
			D_n=\frac{\sum_{i=1}^{n} w_ix_{i}}{\sum_{i=1}^{n} w_i}
			\tag{S3}
			\label{eq:S3}
		\end{equation}
		
		\section{Bare Au@Ag Nanoparticles}
		\label{sec:S2}
		The synthesis procedure of the Au@Ag particles can be found in the main article. After the overgrowth of silver on the spherical gold seeds, the Au@Ag Nanoparticles (NPs) are capped with the surfactant cetyltrimethylammonium chloride (CTAC), the sample was washed once by centrifugation before TEM imaging. The edge length was determined by TEM (Figure\,\ref{fig:S3}\,(a)) as 60$\,\pm\,$1\,nm and electron diffraction proved the single crystalline character of the cubes. Only single crystalline spherical gold seeds result in cubic Au@Ag particles, the yield of 85\% was determined using TEM (Figure\,\ref{fig:S4}).
		\renewcommand{\thefigure}{S3}
		\begin{figure}[h!]
			\centering
			\includegraphics[width=\textwidth]{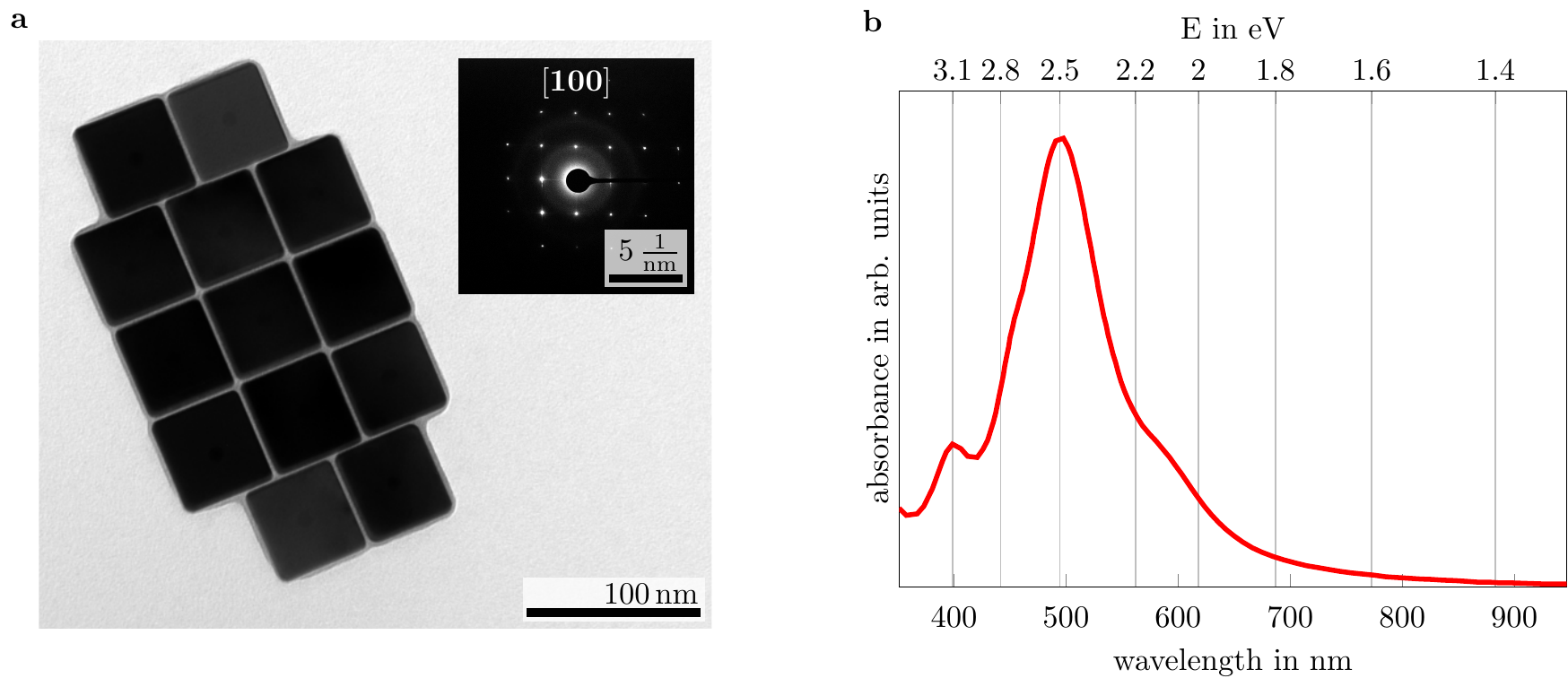}
			\caption{The synthesized Au@Ag NPs were characterized by TEM and UV/Vis. (a) The edge length of the cubes was determined from TEM images. The median edge length was 60$\,\pm\,$1\,nm. Electron diffraction proved the single crystalline character of the cubes. (b) UV/Vis spectroscopy shows an absorption maximum at 495\,nm and a shoulder to higher wavelengths due to impurities (e.g. rods, distorted particles).}
			\label{fig:S3}
		\end{figure}
		\renewcommand{\thefigure}{S4}
		\begin{figure}[h!]
			\centering
			\includegraphics[trim={150 0 0 0}, clip]{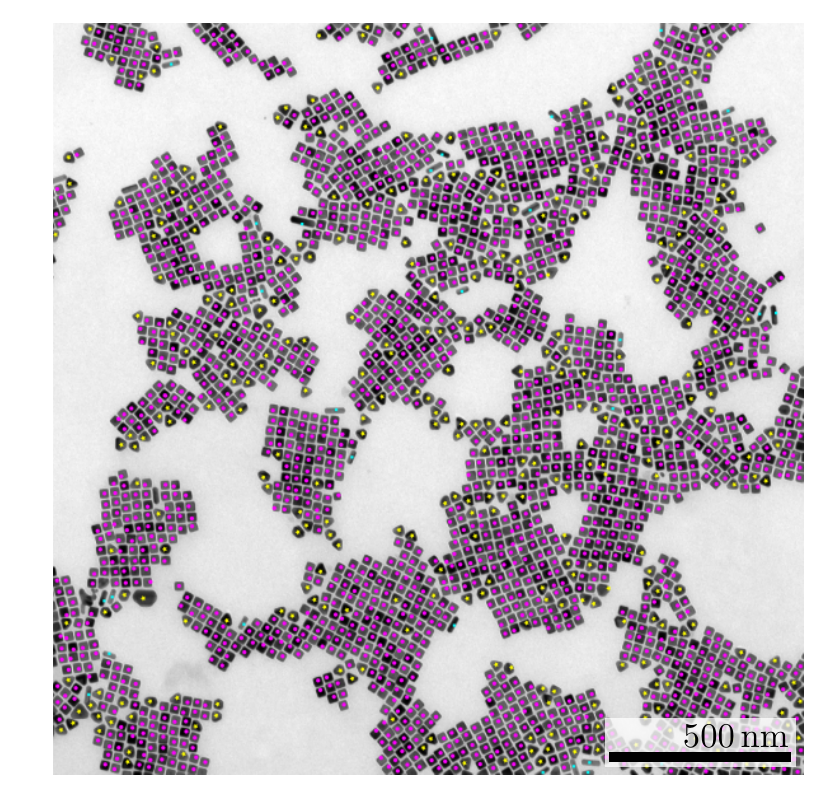}
			\caption{Exemplary set of cubic Au@Ag particles. A synthesis with otherwise same conditions but 4.5\,mL spherical seeds was conducted to yield smaller, more easily countable particles with an edge length of 30\,nm. A yield of 85\,\% cubes was determined by counting 2500 particles.}
			\label{fig:S4}
		\end{figure}
		\newpage
		\section{Silica Shell Overgrowth}
		\label{sec:S3}
		The overgrowth of silica on the Au@Ag nanoparticles requires a transfer to a solution with high contents of ammonia and ethanol. Positively charged nanoparticles (here by CTAC stabilization) are prone to destabilization and agglomeration under these conditions. The polymer can both screen this surface charge and provide steric stabilization. The decrease of surface charge upon functionalization with $\omega$-thiol-terminated polystyrene (PEG-SH) is shown by Zeta-potential measurements (Figure\,\ref{fig:S5}). The synthesis procedure is described in the main article, while reaction parameters and resulting Au@Ag@SiO$_2$ particles can be found in Table\,\ref{tab:S1} and Figure\,\ref{fig:S6}, respectively.
		\renewcommand{\thefigure}{S5}
		\begin{figure}[h!]
			\centering
			\includegraphics[trim={120 0 0 0}, clip]{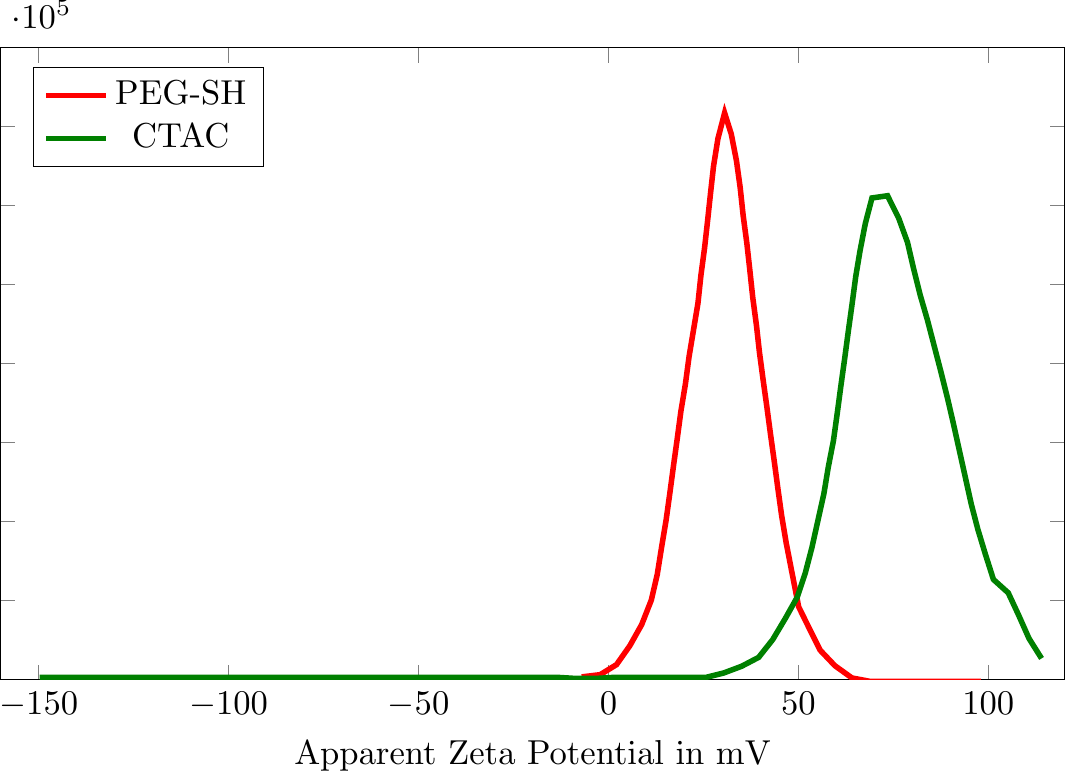}
			\caption{Zeta potential measurements show a decrease of the positive surface charge when the particles where functionalized with PEG-SH. A nanoparticle dispersion with a concentration of 50 µg/mL silver (approx. 1.2$\cdot$10$^{10}$ particles/mL) was measured at 25\,°C with 5$\cdot$12 runs at a count rate of 197.8\,kcps.}
			\label{fig:S5}
		\end{figure}
		\renewcommand{\thefigure}{S6}
		\begin{figure}[h!]
			\centering
			\includegraphics{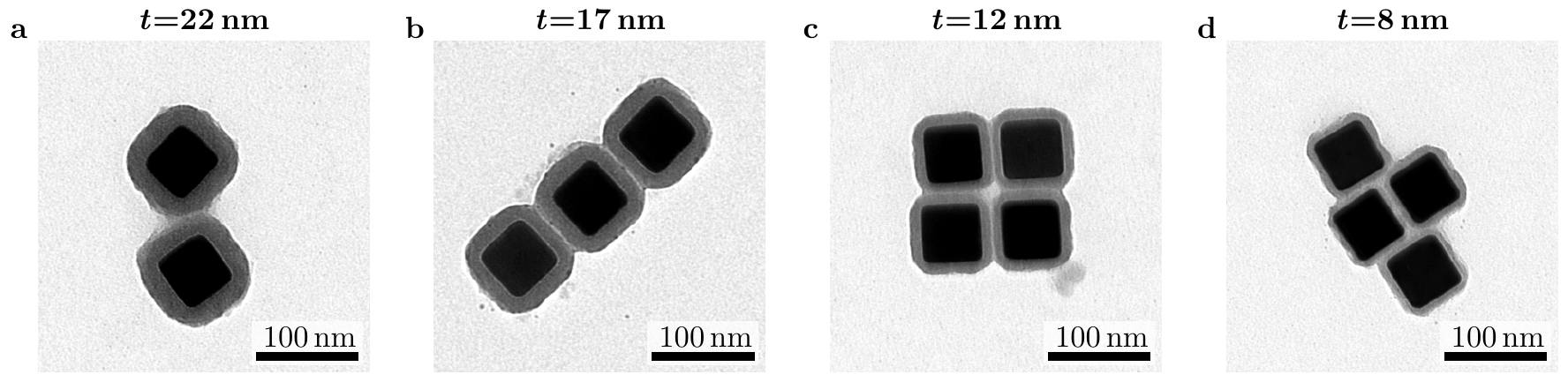}
			\caption{Bright-field TEM images of the investigated Au@Ag@SiO$_2$ particles. Synthesis conditions for the silica shells with a thickness $t$ can be found in Table \ref{tab:S1}.}
			\label{fig:S6}
		\end{figure}
		\renewcommand{\thetable}{S1}
		\begin{table}[h!]
			\caption{Amount of TEOS/EtOH added to the synthesis and the resulting shell thickness. The according particles are depicted in Figure\,\ref{fig:S6}.}
			\label{tab:S1}
			\centering
			\begin{tabular}{|c|c|c|}
				\hline
				TEOS/EtOH volume in $\mu$L
				& Shell thickness $t$ in nm
				& Figure\,\ref{fig:S6} \\
				\hline
				600  & 22  & (a)  \\
				400  & 17  & (b)  \\
				200  & 12  & (c)  \\
				75  & 8  & (d)  \\
				\hline
			\end{tabular}
		\end{table}
		\newpage
		
		\section{Single Particle Localized Surface Plasmon Classification}
		\label{sec:S4}
		To categorize the Localized Surface Plasmon (LSP) modes and understand the observed behavior in more detail, we employ a cubic harmonic basis following atomic orbital / angular momentum classification ($l$$\,=\,$$0\,\widehat{=}\,$s, $l$$\,=\,$$1\,\widehat{=}\,$p, $l$$\,=\,$$2\,\widehat{=}\,$d, $l$$\,=\,$$3\,\widehat{=}\,$f, …). The p, d, and f modes on a cubic NP are shown in Figure\,\ref{fig:S7}\,(a). Note that LSP modes with vanishing angular momentum ($l$=0) violate charge conservation on the NP, thus they are not excited in general. The spatial distribution of the electron energy loss probability for the four modes shown in Figure\,\ref{fig:S7}\,(b) together with the corresponding charge distribution of plane wave excitations (Figure\,\ref{fig:S7}\,(c)) reveal the dominating character of the modes. 
		\renewcommand{\thefigure}{S7}
		\begin{figure}[h!]
			\centering
			\includegraphics[trim={20 0 0 0}, clip]{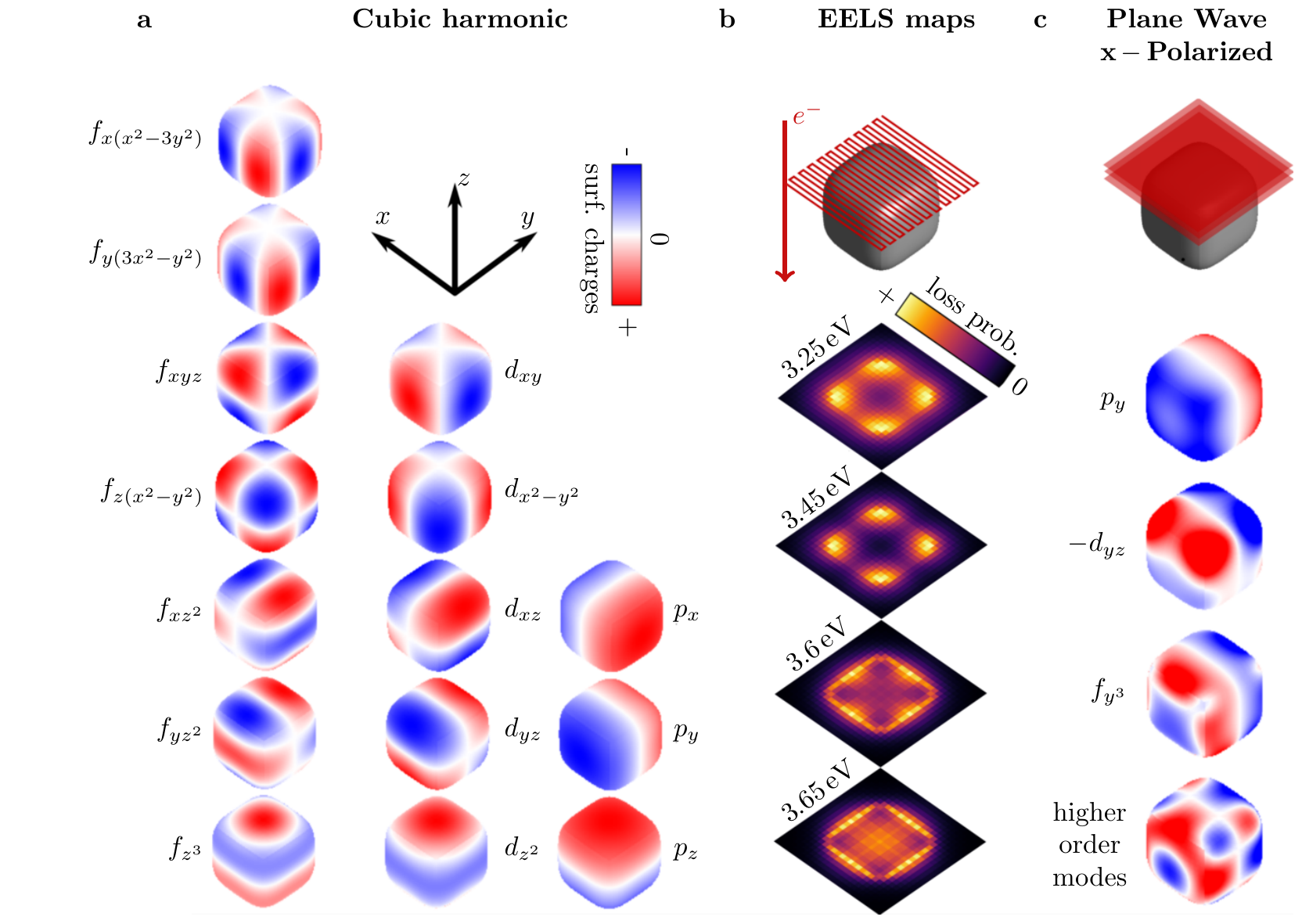}
			\caption{(a) LSP modes on a cubic nanoparticle employing a cubic harmonic basis ($l$$\,=\,$$1\,\widehat{=}\,$p, $l$$\,=\,$$2\,\widehat{=}\,$d, $l$$\,=\,$$3\,\widehat{=}\,$f, …) based on the angular momentum classification of spherical geometries according to Mie theory\cite{Mie1908}. (b) Loss probability maps of four different modes excitable on a 60\,nm silver cube, and (c) corresponding surface charge distributions of plane wave excitation.}
			\label{fig:S7}
		\end{figure}
		\newpage
		\section{Effective Medium Approximation}
		\label{sec:S5}
		In order to understand the excitation energy dependence of the four LSP modes on the silica shell thickness one can study the surface plasmon of a dual layer system of infinite half planes (silica and silver). Whereas one can find an analytically solution for the surface plasmon of a single half plane\cite{Trugler2016}, a solution of the plasmonic response of an arbitrary multilayer system needs numerical support in general. Thus, we study a single half plane and we take into account the silica shell via an effective dielectric media approach. The method to describe complex dielectric surroundings of plasmonic nanoparticles via one effective dielectric function was applied many times for similar problems\cite{Krehl2018, Vernon2010, Chen2012}. We adapted this approach by calculating the effective dielectric medium $\epsilon_{\text{eff}}$ from weighting the dielectric functions of the silica shell and the surrounding medium with the skin depth corresponding to the Beer–Lambert law.\\
		We start by considering a silver-silica interface of an infinitely extended plane. In this case the skin depth can be defined as $|k_{z}^i|^{-1}$ (see e.g.\cite{Trugler2016}). Accordingly, the evanescent plasmon wave in the medium $i$ (silver or silica) is exponentially damped by the factor $e^{-|k_{z}^i||z|}$\cite{Trugler2016, Maier2007} where $z$ and $k_{z}^i$ correspond to the spatial distance and the wavevector perpendicular to the interface. For the infinitely extended silver-silica interface the dispersion relation $k_{z}^i (\omega)$ can be calculated analytically and reads\cite{Trugler2016}
		\begin{equation}
			\tag{S4}
			\label{eq:S4}
			k_{z}^i (\omega)=\frac{\omega}{c}\sqrt{\frac{\epsilon_{i}^2}{\epsilon_0\left(\epsilon_{\text{SiO}_2}+ \epsilon_{\text{Ag}}\right)}}.
		\end{equation}
		
		The calculated value of $k_{z}^i (\omega)$ can be used to model the response of the more complex structure of a silver plane coated by a silica shell of thickness $t$ using the effective dielectric media approach. The effective dielectric function $\epsilon_{\text{eff}}$ can be obtained from averaging the corresponding dielectric functions $(\epsilon_{\text{SiO}_2}$ and $\epsilon_{\text{vac}}=1)$. The weighting factor $w$ of $\epsilon_{\text{SiO}_2}$ can be determined from $k_{z}^{\text{SiO}_2}$ by integrating the normalized intensity along the $z$ axis from 0 to $t$ (blue colored area in Figure\,\ref{fig:S8}\,(a))
		\begin{equation}
			\tag{S5}\label{eq:S5}
			w=\frac{\int_{0}^{t}e^{-z\left|k_z^{\text{SiO}_2}\right|}\text{d}z}{\int_{0}^{\infty}e^{-z\left|k_z^{\text{SiO}_2}\right|}\text{d}z}=1-e^{-t\left|k_z^{\text{SiO}_2}\right|}.
		\end{equation}
		The weighting factor of $\epsilon_{\text{vac}}$ then corresponds to $(1-w)$ as the total area under the blue curve in Figure\,\ref{fig:S8}\,(a) is normalized to 1. For the effective dielectric function follows
		\begin{equation}
			\tag{S6}\label{eq:S6}
			\epsilon_{\text{eff}}=w\epsilon_{\text{SiO}_2}+(1-w)\epsilon_{\text{vac}}.
		\end{equation}
		In Figure\,\ref{fig:S8}\,(c) the resulting effective dielectric functions are plotted for silica shell thicknesses of 8, 22 and 50\,nm, respectively. The larger the skin depth $|k_{z}^i|^{-1}$ plotted in Figure\,\ref{fig:S8}\,(b), the smaller the fraction of $w$ to the total normalized intensity. In consequence the effective dielectric function approaches $\epsilon_{\text{vac}} (\epsilon_{\text{SiO}_2})$ for large (small) values of $|k_{z}^i|^{-1}$.
		\renewcommand{\thefigure}{S8}
		\begin{figure}[h!]
			\centering
			\centering
			\includegraphics{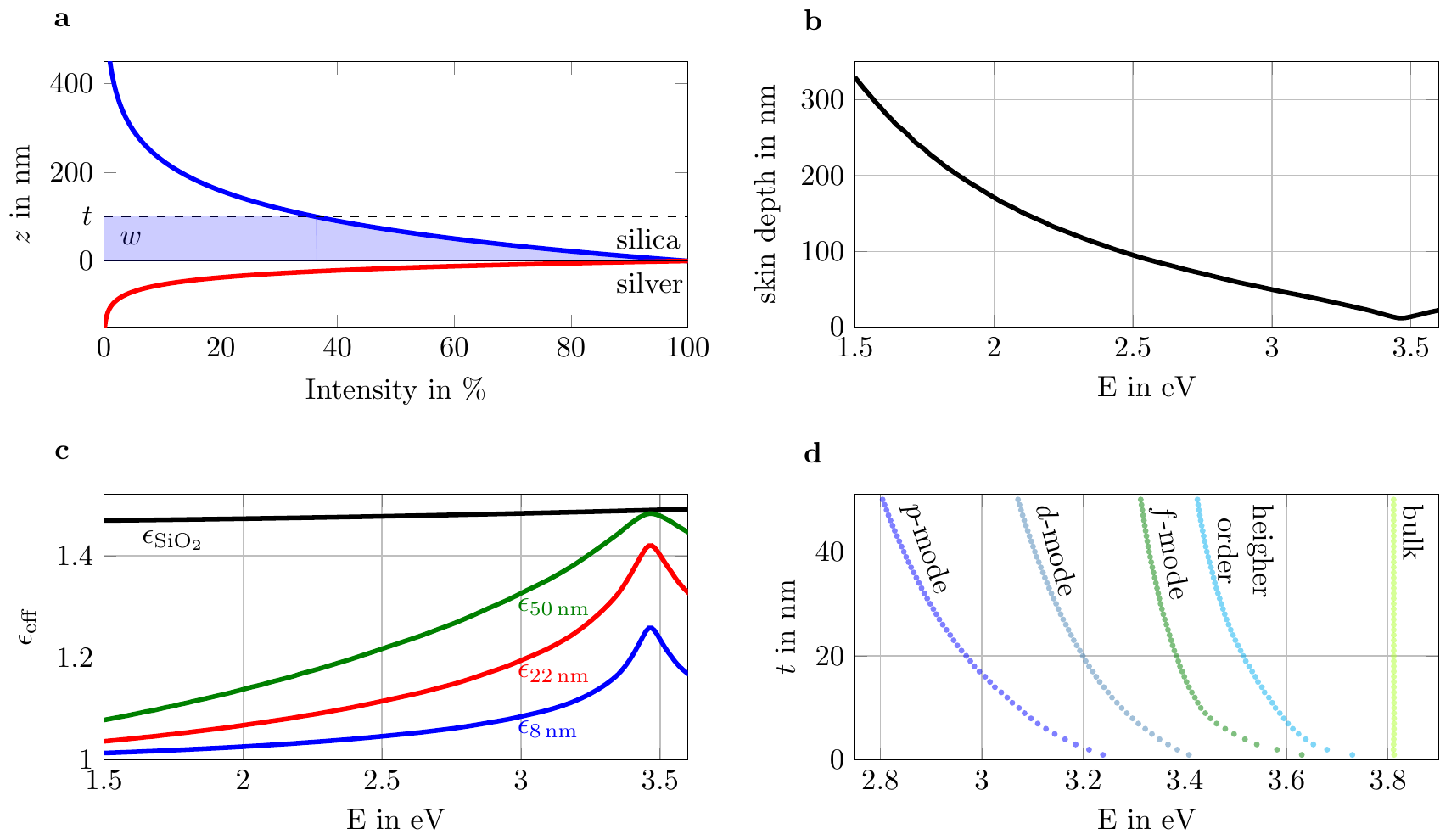}
			\caption{(a) Exponential damping of the Surface Plasmon Polariton (SPP) on a silver-silica interface, in silica (blue graph) and silver (red graph). From the energy dependent skin depth (b) the effective dielectric function $\epsilon_{\text{eff}}$ can be calculated using equation S6. The weighting factor $w$ corresponds to the normalized blue shaded area in (a), the dashed line in (a) indicates the thickness of the silica shell $t$. The resulting effective dielectric functions are plotted in (c) for silica shell thicknesses of 8, 22, and 50\,nm, respectively. The simulated SPP mode energies of a 60\,nm silver cube for effective dielectric functions ranging from $\epsilon_{\text{eff}}(t=1\,\text{nm})$ to $\epsilon_{\text{eff}}(t=50\,\text{nm})$ are plotted in (d).}
			\label{fig:S8}
		\end{figure}
		Using the effective dielectric functions for different values of $t$, the plasmon mode energies of the 60\,nm silver cube were simulated (see Figure\,\ref{fig:S8}\,(d)). The slope of the graphs in Figure\,\ref{fig:S8}\,(d) which correspond to the mode energy shift in dependence of the silica shell thickness $t$ is similar to those of the more realistic two interface simulation (see Figure\,2 in the main text). A quantitative comparison of the effective dielectric function approach with the simulation considering two interfaces is not possible because the value of $k_{z}^i$ depends on the geometry and cannot be calculated analytically for arbitrary shaped NPs. Nevertheless, the concept of an effective dielectric medium allows to understand the behavior of the mode energy shift for different silica shell thicknesses on a qualitative level.
		\newpage
		\section{Whispering Gallery Resonances}
		\label{sec:S6}
		To characterize the Mie resonances at the silica-vacuum interface the response of bare silica cubes of different edge lengths $d$ ranging from 92 to 260\,nm (corresponding to the dimensions of a 60\,nm silver cube with 16 to 100\,nm silica shell) was simulated (Figure\,\ref{fig:S9}). The simulations reveal a redshift of the mode energies as well as an increasing number of excitable modes in the relevant spectral regime with increasing edge length of the cube. Both is expected in case of so called whispering gallery resonances arising from total internal reflection at an interface with dropping dielectric function (e.g., the silica-vacuum interface). Here, the redshift and the increasing number of the resonances can be explained by increasing resonance wavelength of the geometry with increasing size of the NP. Note, however, that we observe a broadening of the typically very sharp resonance peaks of whispering gallery modes in spherical geometries. This may be explained by the cubic shape of the NPs leading to a modified resonance behavior. Due to the above noted properties of the Mie resonances at the silica-vacuum interface we associate them to whispering gallery type modes.
		\renewcommand{\thefigure}{S9}
		\begin{figure}[h!]
			\centering
			\includegraphics{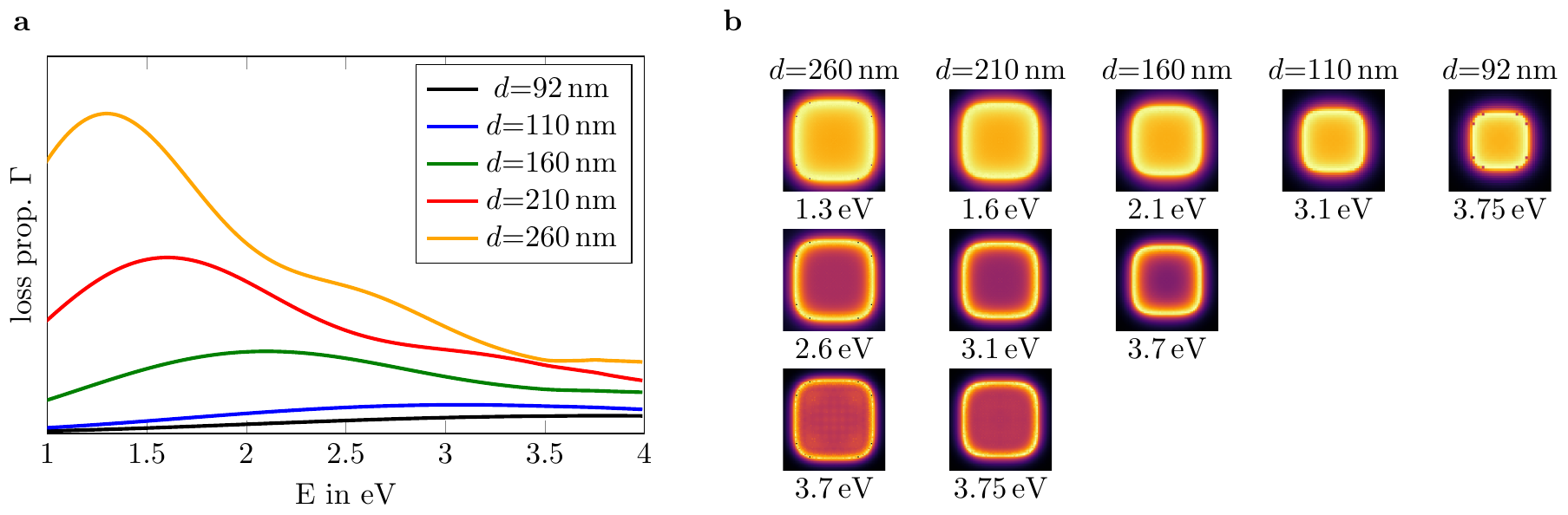}
			\caption{EEL spectra (a) and EEL probability maps of a pure silica cube of different edge length of 92, 110, 160, 210, and 260\,nm (corresponding to the dimensions of a silver cube with 60\,nm edge lengths and a silica shell of 16, 25, 50, 75, and 100\,nm thickness). Due to the dielectric nature of silica $(\epsilon\,$$>$$\,1)$ the excitations may correspond to whispering gallery modes.}
			\label{fig:S9}
		\end{figure}
		\newpage
		\section{Hybridization as Function of Silica Shell Thickness}
		\label{sec:S7}
		In order to proof the enhanced mode hybridization, simulated spectra of bare cubes in vacuum, coated cubes in vacuum and bare cubes embedded in silica (corresponding to a silica shell of infinite with) are compared (Figure\,\ref{fig:S10}). As pointed out in the main text the mode hybridization is determined by the coupling strength of the coupled NPs, where the later can be read-off from the energy splitting of the single particle modes. In case of the uncoated NPs (Figure\,\ref{fig:S10}\,(a),\,(c)) the mode splitting is barely visible for an interparticle gap larger than 48\,nm. On the other hand, in case of the coated NPs (Figure\,\ref{fig:S10}\,(b)) mode splitting is still observable for an interparticle gap of 100\,nm (corresponding to 50\,nm silica shell) due to the amplified mode hybridization.
		\renewcommand{\thefigure}{S10}
		\begin{figure}[h!]
			\centering
			\includegraphics{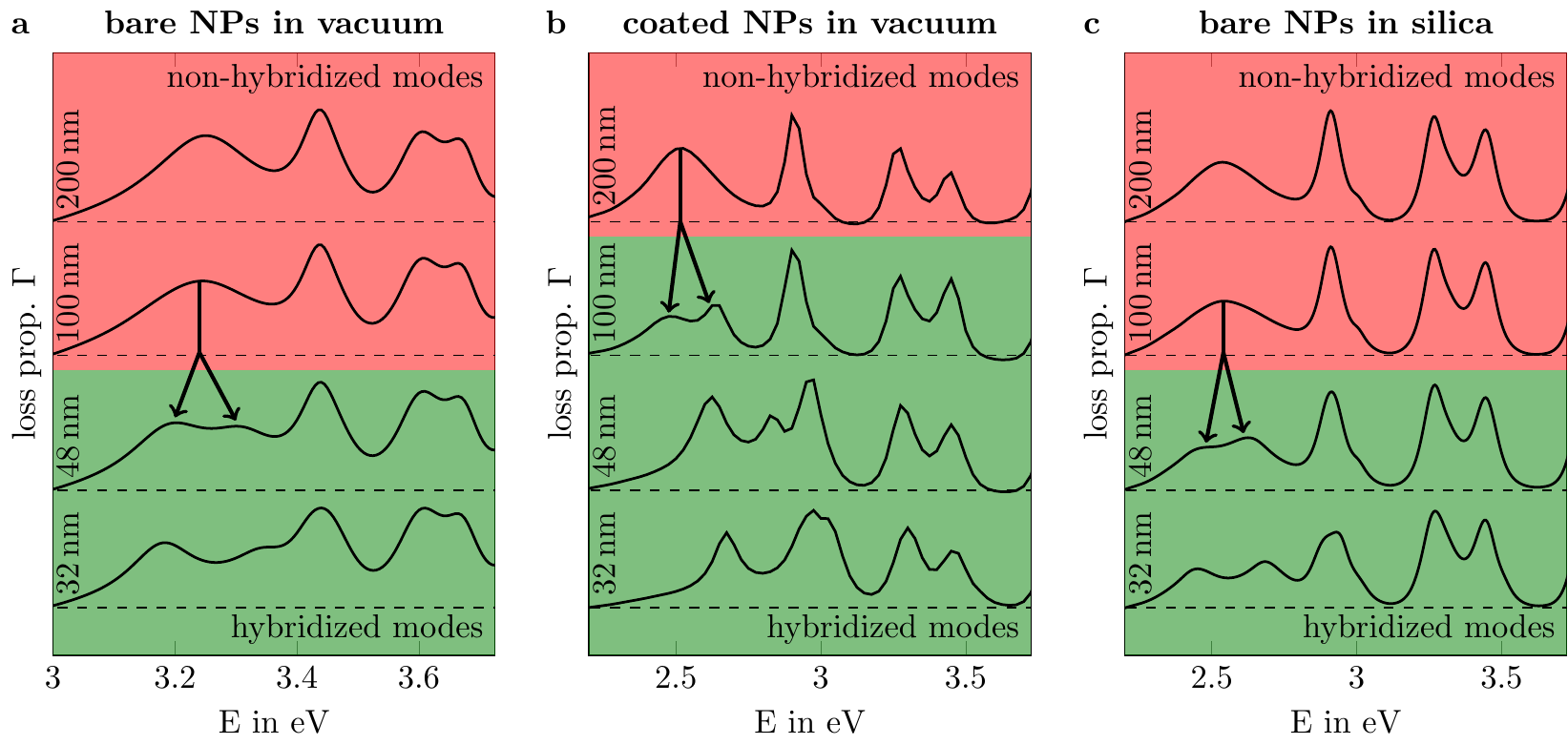}
			\caption{EEL spectra for a silver cube dimer with different interparticle gap (a) embedded in vacuum, (b) silica coated and embedded in vacuum, and (c) embedded in silica. The gap size of the coated dimers in (b) corresponds to the double silica shell thickness, respectively. The spectra reveal a stronger coupling of the coated NPs resulting in mode hybridization for large interparticle gaps of more than 48\,nm.}
			\label{fig:S10}
		\end{figure}

	\medskip
	\bibliographystyle{unsrt}
	\bibliography{plasmon_paper}

\begin{thebibliography}{10}

\bibitem{Anker2008}
Jeffrey~N. Anker, W.~Paige Hall, Olga Lyandres, Nilam~C. Shah, Jing Zhao, and
  Richard~P. {Van Duyne}.
\newblock {Biosensing with plasmonic nanosensors}.
\newblock {\em Nat. Mater.}, 7(6):442--453, 2008.

\bibitem{He2017}
Rui~Xiu He, Robert Liang, Peng Peng, and Y.~{Norman Zhou}.
\newblock {Effect of the size of silver nanoparticles on SERS signal
  enhancement}.
\newblock {\em J. Nanoparticle Res.}, 19(8):267, 2017.

\bibitem{Goris2014}
Bart Goris, Giulio Guzzinati, Cristina Fern{\'{a}}ndez-L{\'{o}}pez, Jorge
  P{\'{e}}rez-Juste, Luis~M. Liz-Marz{\'{a}}n, Andreas Tr{\"{u}}gler, Ulrich
  Hohenester, Jo~Verbeeck, Sara Bals, and Gustaaf {Van Tendeloo}.
\newblock {Plasmon mapping in Au@Ag nanocube assemblies}.
\newblock {\em J. Phys. Chem. C}, 118(28):15356--15362, 2014.

\bibitem{Garcia-Cruz2019}
L~Garcia-Cruz, V~Montiel, and J~Solla-Gullon.
\newblock {Shape-controlled metal nanoparticles for electrocatalytic
  applications}.
\newblock {\em Chem. Nanomater. Vol 1 Met. Nanomater. (Pt a)}, pages 103--156,
  2019.

\bibitem{Schletz2021}
Daniel Schletz, Johannes Schultz, Pavel~L. Potapov, Anja~Maria Steiner, Jonas
  Krehl, Tobias A.~F. König, Martin Mayer, Axel Lubk, and Andreas Fery.
\newblock Exploiting combinatorics to investigate plasmonic properties in
  heterogeneous ag-au nanosphere chain assemblies.
\newblock {\em Advanced Optical Materials}, 9(9):2001983, 2021.

\bibitem{Crespo2014}
Julian Crespo, Andrea Falqui, Jorge Garc{\'{i}}a-Barrasa, Jos{\'{e}}~M.
  L{\'{o}}pez-De-Luzuriaga, Miguel Monge, M.~Elena Olmos, Mar{\'{i}}a
  Rodr{\'{i}}guez-Castillo, Matteo Sestu, and Katerina Soulantica.
\newblock {Synthesis and plasmonic properties of monodisperse Au-Ag alloy
  nanoparticles of different compositions from a single-source organometallic
  precursor}.
\newblock {\em J. Mater. Chem. C}, 2(16):2975--2984, 2014.

\bibitem{Burda2005}
Clemens Burda, Xiaobo Chen, Radha Narayanan, and Mostafa~A. El-Sayed.
\newblock {Chemistry and properties of nanocrystals of different shapes}.
\newblock {\em Chem. Rev.}, 105(4):1025--1102, 2005.

\bibitem{Hanske2018}
Christoph Hanske, Marta~N. Sanz-Ortiz, and Luis~M. Liz-Marz{\'{a}}n.
\newblock {Silica-Coated Plasmonic Metal Nanoparticles in Action}.
\newblock {\em Adv. Mater.}, 30(27):e1707003, 2018.

\bibitem{Vernon2010}
Kristy~C. Vernon, Alison~M. Funston, Carolina Novo, Daniel~E. G{\'{o}}mez, Paul
  Mulvaney, and Timothy~J. Davis.
\newblock {Influence of particle-substrate interaction on localized plasmon
  resonances}.
\newblock {\em Nano Lett.}, 10(6):2080--2086, 2010.

\bibitem{Chen2012}
Huanjun Chen, Feng Wang, Kun Li, Kat~Choi Woo, Jianfang Wang, Quan Li,
  Ling~Dong Sun, Xixiang Zhang, Hai~Qing Lin, and Chun~Hua Yan.
\newblock {Plasmonic percolation: Plasmon-manifested dielectric-to-metal
  transition}.
\newblock {\em ACS Nano}, 6(8):7162--7171, 2012.

\bibitem{Prodan2003}
E.~Prodan, C.~Radloff, N.~J. Halas, and P.~Nordlander.
\newblock A hybridization model for the plasmon response of complex
  nanostructures.
\newblock {\em Science}, 302(5644):419--422, 2003.

\bibitem{Mayer2019}
Martin Mayer, Pavel~L. Potapov, Darius Pohl, Anja~Maria Steiner, Johannes
  Schultz, Bernd Rellinghaus, Axel Lubk, Tobias~A.F. Konig, and Andreas Fery.
\newblock {Direct observation of plasmon band formation and delocalization in
  quasi-infinite nanoparticle chains}.
\newblock {\em Nano Lett.}, 19(6):3854--3862, 2019.

\bibitem{Nordlander2004}
P.~Nordlander, C.~Oubre, E.~Prodan, K.~Li, and M.~I. Stockman.
\newblock {Plasmon hybridization in nanoparticle dimers}.
\newblock {\em Nano Lett.}, 4(5):899--903, 2004.

\bibitem{McMahon2011}
Jeffrey~M. McMahon, Stephen~K. Gray, and George~C. Schatz.
\newblock {Fundamental behavior of electric field enhancements in the gaps
  between closely spaced nanostructures}.
\newblock {\em Phys. Rev. B - Condens. Matter Mater. Phys.}, 83(11):115428,
  2011.

\bibitem{Halas2011}
Naomi~J. Halas, Surbhi Lal, Wei~Shun Chang, Stephan Link, and Peter Nordlander.
\newblock {Plasmons in strongly coupled metallic nanostructures}.
\newblock {\em Chem. Rev.}, 111(6):3913--3961, 2011.

\bibitem{Jain2007}
Prashant~K. Jain, Wenyu Huang, and Mostafa~A. El-Sayed.
\newblock {On the universal scaling behavior of the distance decay of plasmon
  coupling in metal nanoparticle pairs: A plasmon ruler equation}.
\newblock {\em Nano Lett.}, 7(7):2080--2088, 2007.

\bibitem{Hooshmand2019}
Nasrin Hooshmand and Mostafa~A. El-Sayed.
\newblock {Collective multipole oscillations direct the plasmonic coupling at
  the nanojunction interfaces}.
\newblock {\em Proc. Natl. Acad. Sci. U. S. A.}, 116(39):19299--19304, 2019.

\bibitem{Zhu2011}
Wenqi Zhu, Mohamad~G. Banaee, Dongxing Wang, Yizhuo Chu, and Kenneth~B.
  Crozier.
\newblock {Lithographically fabricated optical antennas with gaps well below 10
  nm}.
\newblock {\em Small}, 7(13):1761--1766, 2011.

\bibitem{Menumerov2018}
Eredzhep Menumerov, Spencer~D. Golze, Robert~A. Hughes, and Svetlana Neretina.
\newblock {Arrays of highly complex noble metal nanostructures using
  nanoimprint lithography in combination with liquid-phase epitaxy}.
\newblock {\em Nanoscale}, 10(38):18186--18194, 2018.

\bibitem{Sun2018}
L~Sun, H~Lin, K~L Kohlstedt, G~C Schatz, and C~A Mirkin.
\newblock {Design principles for photonic crystals based on plasmonic
  nanoparticle superlattices}.
\newblock {\em Proc. Natl. Acad. Sci. U. S. A.}, 115(28):7242--7247, 2018.

\bibitem{Gao2012}
Bo~Gao, Gaurav Arya, and Andrea~R. Tao.
\newblock {Self-orienting nanocubes for the assembly of plasmonic
  nanojunctions}.
\newblock {\em Nat. Nanotechnol.}, 7(7):433--437, 2012.

\bibitem{Roller2017}
Eva-Maria Roller, Lucas~V Besteiro, Claudia Pupp, Larousse~Khosravi Khorashad,
  Alexander~O Govorov, and Tim Liedl.
\newblock {Hot spot-mediated non-dissipative and ultrafast plasmon passage}.
\newblock {\em Nat. Phys.}, 13(8):761--765, 2017.

\bibitem{Zhao2020}
Yuan Zhao and Chuanlai Xu.
\newblock {DNA-Based Plasmonic Heterogeneous Nanostructures: Building, Optical
  Responses, and Bioapplications}.
\newblock {\em Adv. Mater.}, 32(41):1907880, 2020.

\bibitem{Bahrig2014}
Lydia Bahrig, Stephen~G. Hickey, and Alexander Eychmüller.
\newblock Mesocrystalline materials and the involvement of oriented attachment
  – a review.
\newblock {\em CrystEngComm}, 16:9408--9424, 2014.

\bibitem{Brunner2020}
Julian Brunner, Britta ~, Rose Rosenberg, Sebastian Sturm, Helmut C{\"{o}}lfen,
  and Elena~V. Sturm.
\newblock {Nonclassical Recrystallization}.
\newblock {\em Chem. - A Eur. J.}, 26(66):15242--15248, 2020.

\bibitem{Kapuscinski2020}
Martin Kapuscinski, Pierre Munier, Mo~Segad, and Lennart Bergstr{\"{o}}m.
\newblock {Two-Stage Assembly of Mesocrystal Fibers with Tunable Diameters in
  Weak Magnetic Fields}.
\newblock {\em Nano Lett.}, 20(10):7359--7366, 2020.

\bibitem{Sturm2016}
Elena~V. Sturm~(née Rosseeva) and Helmut Cölfen.
\newblock Mesocrystals: structural and morphogenetic aspects.
\newblock {\em Chem. Soc. Rev.}, 45:5821--5833, 2016.

\bibitem{Jana2001}
N.~R. Jana, L.~Gearheart, and C.~J. Murphy.
\newblock {Seed-mediated growth approach for shape-controlled synthesis of
  spheroidal and rod-like gold nanoparticles using a surfactant template}.
\newblock {\em Adv. Mater.}, 13(18):1389--1393, 2001.

\bibitem{Niu2013}
Wenxin Niu, Ling Zhang, and Guobao Xu.
\newblock {Seed-mediated growth of noble metal nanocrystals: Crystal growth and
  shape control}.
\newblock {\em Nanoscale}, 5(8):3172--3181, 2013.

\bibitem{Heinz2017}
Hendrik Heinz, Chandrani Pramanik, Ozge Heinz, Yifu Ding, Ratan~K. Mishra,
  Delphine Marchon, Robert~J. Flatt, Irina Estrela-Lopis, Jordi Llop, Sergio
  Moya, and Ronald~F. Ziolo.
\newblock {Nanoparticle decoration with surfactants: Molecular interactions,
  assembly, and applications}.
\newblock {\em Surf. Sci. Rep.}, 72(1):1--58, 2017.

\bibitem{Kirner2020}
Felizitas Kirner, Pavel Potapov, Johannes Schultz, Jessica Geppert, Magdalena
  M{\"{u}}ller, Guillermo Gonz{\'{a}}lez-Rubio, Sebastian Sturm, Axel Lubk, and
  Elena Sturm.
\newblock {Additive-controlled synthesis of monodisperse single crystalline
  gold nanoparticles: Interplay of shape and surface plasmon resonance}.
\newblock {\em J. Mater. Chem. C}, 8(31):10844--10851, 2020.

\bibitem{Grzelczak2008}
Marek Grzelczak, Jorge P{\'{e}}rez-Juste, Paul Mulvaney, and Luis~M.
  Liz-Marz{\'{a}}n.
\newblock {Shape control in gold nanoparticle synthesis}.
\newblock {\em Chem. Soc. Rev.}, 37(9):1783--1791, 2008.

\bibitem{Boles2016}
Michael~A. Boles, Michael Engel, and Dmitri~V. Talapin.
\newblock {Self-assembly of colloidal nanocrystals: From intricate structures
  to functional materials}.
\newblock {\em Chem. Rev.}, 116(18):11220--11289, 2016.

\bibitem{Zheng2013}
Yiqun Zheng, Yanyun Ma, Jie Zeng, Xiaolan Zhong, Mingshang Jin, Zhi~Yuan Li,
  and Younan Xia.
\newblock {Seed-mediated synthesis of single-crystal gold nanospheres with
  controlled diameters in the range 5-30 nm and their self-assembly upon
  dilution}.
\newblock {\em Chem. - An Asian J.}, 8(4):792--799, 2013.

\bibitem{Jeon2014}
Seog-Jin Jeon, Jae-Hwang Lee, and Edwin~L. Thomas.
\newblock Polyol synthesis of silver nanocubes via moderate control of the
  reaction atmosphere.
\newblock {\em Journal of Colloid and Interface Science}, 435:105--111, 2014.

\bibitem{Gomez-Grana2013}
Sergio G{\'{o}}mez-Gra{\~{n}}a, Bart Goris, Thomas Altantzis, Cristina
  Fern{\'{a}}ndez-L{\'{o}}pez, Enrique Carb{\'{o}}-Argibay, Andr{\'{e}}s
  Guerrero-Mart{\'{i}}nez, Neyvis Almora-Barrios, Nuria L{\'{o}}pez, Isabel
  Pastoriza-Santos, Jorge P{\'{e}}rez-Juste, Sara Bals, Gustaaf {Van Tendeloo},
  and Luis~M. Liz-Marz{\'{a}}n.
\newblock {Au@Ag nanoparticles: Halides stabilize {\{}100{\}} facets}.
\newblock {\em J. Phys. Chem. Lett.}, 4(13):2209--2216, 2013.

\bibitem{Sotiriou2010}
Georgios~A. Sotiriou, Takumi Sannomiya, Alexandra Teleki, Frank Krumeich, Janos
  V{\"{o}}r{\"{o}}s, and Sotiris~E. Pratsinis.
\newblock {Non-toxic dry-coated nanosilver for plasmonic biosensors}.
\newblock {\em Adv. Funct. Mater.}, 20(24):4250--4257, 2010.

\bibitem{Stober1968}
Werner St{\"{o}}ber, Arthur Fink, and Ernst Bohn.
\newblock {Controlled growth of monodisperse silica spheres in the micron size
  range}.
\newblock {\em J. Colloid Interface Sci.}, 26(1):62--69, 1968.

\bibitem{Kobayashi2005}
Yoshio Kobayashi, Hironori Katakami, Eiichi Mine, Daisuke Nagao, Mikio Konno,
  and Luis~M. Liz-Marzán.
\newblock Silica coating of silver nanoparticles using a modified stöber
  method.
\newblock {\em Journal of Colloid and Interface Science}, 283(2):392--396,
  2005.

\bibitem{Pastoriza-Santos2006}
Isabel Pastoriza-Santos, Jorge P{\'{e}}rez-Juste, and Luis~M. Liz-Marz{\'{a}}n.
\newblock {Silica-coating and hydrophobation of CTAB-stabilized gold nanorods}.
\newblock {\em Chem. Mater.}, 18(10):2465--2467, 2006.

\bibitem{Hung2013}
L.~Hung, S.~Y. Lee, O.~McGovern, O.~Rabin, and I.~Mayergoyz.
\newblock {Calculation and measurement of radiation corrections for plasmon
  resonances in nanoparticles}.
\newblock {\em Phys. Rev. B - Condens. Matter Mater. Phys.}, 88(7):75424, 2013.

\bibitem{Rodriguez-Fernandez2007}
Jessica Rodr{\'{i}}guez-Fern{\'{a}}ndez, Isabel Pastoriza-Santos, Jorge
  P{\'{e}}rez-Juste, F.~Javier {Garc{\'{i}}a De Abajo}, and Luis~M.
  Liz-Marz{\'{a}}n.
\newblock {The effect of silica coating on the optical response of
  sub-micrometer gold spheres}.
\newblock {\em J. Phys. Chem. C}, 111(36):13361--13366, 2007.

\bibitem{Kluczyk-Korch2019}
Katarzyna Kluczyk-Korch, Christin David, Witold Jacak, and Janusz Jacak.
\newblock {Application of core-shell metallic nanoparticles in hybridized
  perovskite solar cell-various channels of plasmon photovoltaic effect}.
\newblock {\em Materials (Basel).}, 12(19):3192, 2019.

\bibitem{Montano-Priede2017}
Jos{\'{e}}~Luis Monta{\~{n}}o-Priede, Jo{\~{a}}o~Paulo Coelho, Andr{\'{e}}s
  Guerrero-Mart{\'{i}}nez, Ovidio Pe{\~{n}}a-Rodr{\'{i}}guez, and Umapada Pal.
\newblock {Fabrication of Monodispersed Au@SiO2Nanoparticles with Highly Stable
  Silica Layers by Ultrasound-Assisted St{\"{o}}ber Method}.
\newblock {\em J. Phys. Chem. C}, 121(17):9543--9551, 2017.

\bibitem{Nelayah2007}
Jaysen Nelayah, Mathieu Kociak, Odile St{\'{e}}phan, F.~Javier~Garc{\'{i}}a {De
  Abajo}, Marcel Tenc{\'{e}}, Luc Henrard, Dario Taverna, Isabel
  Pastoriza-Santos, Luis~M. Liz-Marz{\'{a}}n, and Christian Colliex.
\newblock {Mapping surface plasmons on a single metallic nanoparticle}.
\newblock {\em Nat. Phys.}, 3(5):348--353, 2007.

\bibitem{Schmidt2012}
Franz-Philipp Schmidt, Harald Ditlbacher, Ulrich Hohenester, Andreas Hohenau,
  Ferdinand Hofer, and Joachim~R. Krenn.
\newblock Dark plasmonic breathing modes in silver nanodisks.
\newblock {\em Nano Letters}, 12(11):5780--5783, Nov 2012.

\bibitem{Martin2014}
Jérôme Martin, Mathieu Kociak, Zackaria Mahfoud, Julien Proust, Davy Gérard,
  and Jérôme Plain.
\newblock High-resolution imaging and spectroscopy of multipolar plasmonic
  resonances in aluminum nanoantennas.
\newblock {\em Nano Letters}, 14(10):5517--5523, 2014.

\bibitem{Yoshimoto2018}
Daichi Yoshimoto, Hikaru Saito, Satoshi Hata, Yoshifumi Fujiyoshi, and Hiroki
  Kurata.
\newblock Characterization of nonradiative bloch modes in a plasmonic
  triangular lattice by electron energy-loss spectroscopy.
\newblock {\em ACS Photonics}, 5(11):4476--4483, 2018.

\bibitem{Hohenester2012}
Ulrich Hohenester and Andreas Tr{\"{u}}gler.
\newblock {MNPBEM - A Matlab toolbox for the simulation of plasmonic
  nanoparticles}.
\newblock {\em Comput. Phys. Commun.}, 183(2):370--381, 2012.

\bibitem{Hohenester2014}
Ulrich Hohenester.
\newblock {Simulating electron energy loss spectroscopy with the MNPBEM
  toolbox}.
\newblock {\em Comput. Phys. Commun.}, 185(3):1177--1187, 2014.

\bibitem{Yang2015}
Honghua~U. Yang, Jeffrey D'Archangel, Michael~L. Sundheimer, Eric Tucker,
  Glenn~D. Boreman, and Markus~B. Raschke.
\newblock {Optical dielectric function of silver}.
\newblock {\em Phys. Rev. B - Condens. Matter Mater. Phys.}, 91(23):235137,
  2015.

\bibitem{Gao2012a}
Lihong Gao, Fabien Lemarchand, and Michel Lequime.
\newblock {Exploitation of multiple incidences spectrometric measurements for
  thin film reverse engineering}.
\newblock {\em Opt. Express}, 20(14):15734, 2012.

\bibitem{Schindelin2012}
Johannes Schindelin, Ignacio Arganda-Carreras, Erwin Frise, Verena Kaynig, Mark
  Longair, Tobias Pietzsch, Stephan Preibisch, Curtis Rueden, Stephan Saalfeld,
  Benjamin Schmid, Jean~Yves Tinevez, Daniel~James White, Volker Hartenstein,
  Kevin Eliceiri, Pavel Tomancak, and Albert Cardona.
\newblock {Fiji: An open-source platform for biological-image analysis}.
\newblock {\em Nat. Methods}, 9(7):676--682, 2012.

\bibitem{Mie1908}
Gustav Mie.
\newblock {Beitr{\"{a}}ge zur Optik tr{\"{u}}ber Medien, speziell kolloidaler
  Metall{\"{o}}sungen}.
\newblock {\em Ann. Phys.}, 330(3):377--445, 1908.

\bibitem{Trugler2016}
Andreas Tr{\"{u}}gler.
\newblock {\em {Optical Properties of Metallic Nanoparticles : Basic Principles
  and Simulation}}, volume 232.
\newblock Springer, 2016.

\bibitem{Krehl2018}
J.~Krehl, G.~Guzzinati, J.~Schultz, P.~Potapov, D.~Pohl, J{\'{e}}r{\^{o}}me
  Martin, J.~Verbeeck, A.~Fery, B.~B{\"{u}}chner, and A.~Lubk.
\newblock {Spectral field mapping in plasmonic nanostructures with nanometer
  resolution}.
\newblock {\em Nat. Commun.}, 9(1):1--6, 2018.

\bibitem{Maier2007}
Stefan~Alexander Maier.
\newblock {\em {Plasmonics: fundamentals and applications}}.
\newblock Springer Science {\&} Business Media, 2007.

\end{thebibliography}

\end{document}